\documentclass{aa}

\usepackage{graphicx}
\usepackage{float}
\DeclareGraphicsExtensions{.pdf}
\usepackage[T1]{fontenc}
\usepackage[utf8]{inputenc} 
\usepackage{bm}
\usepackage[english]{babel} 
\usepackage{acronym}
\usepackage{xspace}
\usepackage{footmisc}
\usepackage[hidelinks,colorlinks=true]{hyperref}
\usepackage{lineno}
\usepackage{extdash}
\usepackage{makecell}
\usepackage{xcolor}
\usepackage{hyperref}
\hypersetup{breaklinks=true}

\newcommand{\hessj}{HESS$\,$J1813$-$178}
\newcommand{\msh}{MSH$\,$15$-$52}

\begin{document}

\title{High-Resolution Measurements with the CTAO Southern Array: The Case for Pulsar Wind Nebulae} 

\titlerunning{High-Resolution Measurements with the CTAO Southern Array: The Case for PWNe}

\author{G. Schwefer
          \and
          J. A. Hinton 
          }

   \institute{Max-Planck-Institut f\"ur Kernphysik, Saupfercheckweg 1, 69117 Heidelberg\\
              \email{georg.schwefer@mpi-hd.mpg.de}
             }

\authorrunning{G. Schwefer \& J. A. Hinton
          }
          
 
  \abstract
   {The advent of the Cherenkov Telescope Array Observatory (CTAO) and recent advances in reconstruction of gamma-ray photons with Cherenkov telescopes are bound to push the limit of angular resolution to an - in the gamma-ray band - unprecedented precision of less than one arcminute at tens of TeV. Naturally, such instrumental improvements open up possibilities for new and interesting scientific studies.}
   {We aim to show that the study of pulsar wind nebulae (PWNe) in particular is bound to profit from these high-resolution measurements. This is because PWNe are the dominant Galactic source population at TeV energies, exhibit hard spectra up to hundreds of TeV and from X-ray observations are known to possess plentiful structure on arcminute scales. }
   {Using \hessj\ and \msh\ as examples, we create simple leptonic models of the TeV morphology of these sources based on X-ray observations and existing gamma-ray measurements. Then, assuming different models for the exposure and point spread function of the observatory, we create mock observations with the future CTAO southern array. We use these to assess the ability of these observations to differentiate between models and study the physics of these sources, in particular to infer the structure of the magnetic field and electron distributions.}
   {We find that future observations with the CTAO southern array at multi-TeV energies -- in combination with existing X-ray measurements -- will likely be able to constrain the distributions of magnetic field and high-energy electrons in these sources. We demonstrate that the sensitivity of these measurements can be significantly enhanced with the improved angular resolution achievable with novel reconstruction algorithms. However, we also show that in the relevant multi-TeV regime, signal-photon statistics remain a limitation and trading event statistics for improved angular resolution is not necessarily advantageous.}
   {}

\maketitle

\section{Introduction\label{sec:introduction}}

Astronomical observations at any wavelength shorter than the X-ray have always suffered from comparatively poor angular resolution. Satellite-based gamma-ray instruments and ground-based TeV instruments currently reach resolutions of at best a few arcminutes (see e.g.~\cite{Fermi-LAT:2019yla,Parsons:2014voa}). The advent of the Cherenkov Telescope Array Observatory (CTAO)~\citep{CTAConsortium:2017dvg}, coupled with cutting-edge angular reconstruction techniques~\citep{Schwefer:2024bsx}, promise a very significant advance in achievable angular resolution.

In particular, this applies to the foreseen southern array of the CTAO. In its \emph{Alpha}-configuration, it will feature 37 small-sized telescopes (SSTs) with a primary mirror diameter of $\approx4\,\mathrm{m}$ spread over an unprecdented area of $\approx 3\,\mathrm{km}^2$~\citep{CTAConsortium:2017dvg,Trois:2024nbx}. As such, the array is optimised for observations in the multi-TeV energy range and it appears possible to achieve an angular resolution well below $1$ arcminute at energies above $10\,\mathrm{TeV}$~\citep{Schwefer:2024bsx}. This is a great improvement over the resolution of a couple of arcminutes possible with current-generation instruments such as H.E.S.S.~\citep{Parsons:2014voa}.

Naturally, these prospects raise the question of which studies of objects and physical processes in the gamma-ray domain can profit most from this improved angular resolution.

Suitable targets need to satisfy two criteria: First, they need to be sufficiently bright in the multi-TeV energy range where the best angular resolution is available. In this range, observations are typically limited by the effective area of the observatory and thus by the statistics of the measured number of photons. The second criterion is that there must be resolvable structure and corresponding physical insights to be obtained on this arcminute scale in the sources.

Pulsar Wind Nebulae (PWNe) satisfy both criteria:
On one hand, they are the dominating source class in the Galactic plane at multi-TeV energies. For instance, a large fraction of the sources identified in the H.E.S.S. Galactic plane survey~\citep{HESS:2018pbp} and the recent first LHAASO catalogue~\citep{LHAASO:2023rpg} are associated with PWNe. Whilst more power is thought to go in to acceleration in supernova remnants (SNRs), electron/positron pairs accelerated in PWN are both very efficiently accelerated and also very efficient in the production of TeV inverse Compton emission, provided they can escape very high magnetic field regions close to young pulsars (see e.g.~\cite{Breuhaus:2020mof}). On the other hand, PWNe are also known to possess considerable structure on arcminute scales. This knowledge stems primarily from X-ray observations of the synchrotron emission from (in particular) the high-field regions in the PWNe with observatories such as Chandra~\citep{Weisskopf:2001uu}, XMM-Newton~\citep{XMM:2001haf} and now~eROSITA~\citep{eROSITA:2020emt}. 

Here, we identify two objects, \hessj\ and \msh, which we use to demonstrate the scientific promise of high-resolution observations of  the inverse Compton emission from PWNe at multi-TeV energies highlighting the connection to the corresponding synchrotron emission in the X-ray band. These two objects are chosen based on their relative brightness, their location in the sky (accessible to the southern CTA Observatory site), and the presence of X-ray structures on the relevant scales.

\hessj\ is located at $(\alpha=273.40^\circ,\;\delta=-17.83^\circ)$~\citep{HESS:2024usd} and was discovered first as a compact gamma-ray source of unidentified type~\citep{2005Sci...307.1938A}. An X-ray counterpart was identified shortly thereafter~\citep{Ubertini:2005nh,Brogan:2005hg} and characterised as a PWN through XMM-Newton~\citep{Funk:2006xk} and Chandra~\citep{Helfand:2007ys} observations. The X-ray nebula is observed to possess a distinct asymmetrical structure that is most likely a consequence of the interaction with the SNR shell visible in radio observations~\citep{Brogan:2005hg, Funk:2006xk}. The pulsed emission from the neutron star powering the system was later identified in X-rays~\citep{Gotthelf:2009ph,Halpern:2012rs} and recently also in the radio band~\citep{Camilo:2021jmy}. With a spin-down luminosity of $\dot{E}= 5.6\times 10^{37}\,\mathrm{erg}\mathrm{s}^{-1}$, it is the second-most powerful known pulsar in the Milky Way~\citep{Camilo:2021jmy}. Its characteristic age is $t_{\mathrm{c}}=5600\,\mathrm{yrs}$~\citep{Camilo:2021jmy}. Its distance is unknown, the current best estimates from the dispersion measure of the pulsar are either $6.2\,\mathrm{kpc}$ or $12\,\mathrm{kpc}$~\citep{Camilo:2021jmy}. A recent study of the region combining data from H.E.S.S. and Fermi-LAT~\citep{HESS:2024usd} found the region to contain two distinct sources: The known compact source associated with the PWN and an extended gamma-ray source around the PWN on the scale of one degree. The emission from this extended source has been attributed to inverse Compton emission from escaped electrons and positrons from the PWN.

Like \hessj, the system \msh\ is a composite SNR-PWN system. At its heart at a location of $(\alpha=228.48^\circ,\;\delta=-59.14^\circ)$ sits the pulsar PSR$\,$B1509$-$58. With a spin-down luminosity of $\dot{E}= 1.8\times 10^{37}\,\mathrm{erg}\,\mathrm{s}^{-1}$ it is also among the most powerful known pulsars in the Milky Way~\citep{Gaensler:1999xf}. The estimated characteristic age is $t_{\mathrm{c}}=1700\,\mathrm{yrs}$~\citep{Gaensler:1999xf}. In the same study, the distance to the pulsar is estimated as $5.2\pm1.4\,\mathrm{kpc}$.

In X-rays, the system has been extensively observed using e.g. Chandra~\citep{Gaensler:2001ac}, NuStar~\citep{An:2014haa} and XMM-Newton~\citep{Schock:2010fq}. These observations have revealed several components: Besides the pulsar itself, there is elongated non-thermal emission from the PWN and a region of thermal emission to the north of the pulsar. This thermal emission is coincident with the RCW 89 emission nebula seen in radio and associated with the SNR shell. The PWN itself contains further substructure. Most notably, there is a jet extending to the south-east of the pulsar and a number of filaments to the north and west of the pulsar connecting to the thermal emission (reminiscent of fingers, hence the colloquial label of "cosmic hand" for this system). See~\cite{Gaensler:2001ac} for a more in-depth discussion of these features.
Most recently, IXPE also studied the polarisation of the X-rays~\citep{Romani:2023zoh} and found the magnetic field to align with the structures in the source.

In TeV gamma-rays, \msh\ was first detected and studied by H.E.S.S.~\citep{HESS:2005lqd} where an elongated extended source was found coincident with the X-ray nebula. The source is found to possess a hard spectrum up to at least $50\,\mathrm{TeV}$.

More recently, using an approach similar to the one taken in this paper, the morphology of the TeV source was studied in more detail using H.E.S.S. data and Chandra X-ray observations~\citep{Tsirou:2017itf,Tsirou:2019kbd}. Whilst some constraints on the magnetic field structure could be derived, the studies are limited by the angular resolution of H.E.S.S..

In this paper, we show how high-energy and high-resolution measurements with the future CTAO Southern array could aid in revealing the properties of these PWNe, in particular the spatial structure of the magnetic field and that of relativistic electrons and positrons.

In Section~\ref{sec:approach}, we describe our approach to generate simple morphological models of \hessj\ and \msh\ based on X-ray observations, the instrument response functions (IRFs) we use to create mock observations of these sources, and the statistical method used to evaluate the power to differentiate between the models. We present the results of our study including the validation of our models in Section~\ref{sec:results} and summarise and conclude in Section~\ref{sec:conclusion}.

\section{Approach}
\label{sec:approach}
\subsection{Source Modeling}
\label{sec:source_modelling}
The goal of the source modeling is to create different hypotheses for the multi-TeV morphology of \hessj\ and \msh\ based on the spatially resolved X-ray measurements of the sources. We do not aim to provide a detailed model of the PWN that for example takes into account the evolutionary history of the pulsar or its progenitor~\citep[see e.g.][]{refId1}, specific particle transport processes for high-energy particles or magnetic field geometries (see e.g.~\cite{Volpi:2008ng} for a model of the radiation from a PWN based directly on MHD simulations), but rather a snapshot model focusing on the current properties and indicative of what could be revealed by CTAO. 

We make the standard assumption that the non-thermal X-ray emission is produced via synchrotron radiation from high-energy electrons and that the TeV gamma-rays originate from inverse Compton emission from the same high-energy electrons scattering with soft target photons. We can then make use of the following scaling relations:
\begin{equation}
    I_{\mathrm{X}}\propto n_{\mathrm{e}}\times B^2
\end{equation}
and
\begin{equation}
\label{eq:TeV_radiation}
    I_{\mathrm{TeV}}\propto n_{\mathrm{e}}\times n_{\mathrm{rad}}
\end{equation}

In this, $n_{\mathrm{e}}$ is the density of high-energy electrons, $B$ the magnetic field strength and $n_{\mathrm{rad}}$ the intensity of the target radiation field.

Generally, the target radiation field for PWNe comprises of the Interstellar radiation field (ISRF), the Cosmic Microwave background (CMB) and the synchrotron radiation present in the sources~\citep{2017ASSL..446..161G}. The latter and the associated synchrotron self-Compton mechanism can play an important role in particularly young, compact and powerful PWNe such as the Crab nebula~\citep{Dirson:2022lnl}, but is subdominant in the vast majority of sources~\citep{Torres:2013jha}.

Under the assumption that the target radiation field is dominated by the ISRF and CMB and therefore approximately homogeneous on the scale of the source considered, the TeV morphology of the sources directly corresponds to the distribution of high-energy electrons $n_{\mathrm{e}}$.

Different hypotheses for $n_{\mathrm{e}}$ can be generated from the measured X-ray intensity by dividing it by different models for the magnetic field in the source

\begin{equation}
\label{eq:TeV_from_Xray}
    I_{\mathrm{TeV}}\propto n_{\mathrm{e}}\propto \frac{I_{\mathrm{X}}} {B^2}.
\end{equation}

This, in turn, means that resolving the TeV morphology directly gives insights into both the distribution of high-energy electrons and the magnetic field present in the sources and lifts the degeneracy present in the X-ray observations.

As mentioned above, this modeling approach is based on the assumption that the same (energy range of) electrons that produce the X-rays also produce the TeV gamma rays. This depends on both the strength of the magnetic field and the spectrum of the target radiation field, i.e. the CMB and ISRF for the case of PWNe. 
For a typical magnetic field strength of $10\,\mathrm{\mu G}$ (e.g.~\cite{HESS:2024usd,Gaensler:2001ac}), electrons with energies between around $50\,\mathrm{TeV}$ and $100\,\mathrm{TeV}$ typically produce X-rays in the $2-9\,\mathrm{keV}$ band~\citep{Hinton:2009zz,Blumenthal:1970gc}. Taking into account the non-negligible Klein-Nishina effects, the peaks of the gamma-ray SEDs produced in inverse Compton scattering by these same electrons lie between $10\,\mathrm{TeV}$ and $40\,\mathrm{TeV}$ for scattering of the CMB and between $45\,\mathrm{TeV}$ and $95\,\mathrm{TeV}$ for the scattering of far-infrared photons ($E_{\mathrm{FIR}}=0.02\,\mathrm{eV}$)~\citep{Hinton:2009zz,Blumenthal:1970gc}.

Therefore, the $10-100\,\mathrm{TeV}$ energy range can be considered well matched with the $2-10\,\mathrm{keV}$ X-ray band and we restrict our modeling and analysis to this energy range and assume an energy-independent morphology within it.

These considerations also show that harnessing the synergies between X-ray and gamma-ray measurements requires a focus on multi-TeV energies, towards the upper end of the sensitive energy range of current IACTs. The fact that this is also the energy range where the angular resolution of IACTs is best provides a unique opportunity to perform morphological studies such as the one proposed here and study the structure of these sources.

In practice, our morphological modeling consists of two steps: First, we generate models of the morphology of the $2-10\,\mathrm{keV}$ X-ray emission for both sources. Then, using Equation~\ref{eq:TeV_from_Xray}, we generate a set of models for the corresponding $10-100\,\mathrm{TeV}$ gamma-ray morphology.

In the following, we first describe our modeling of the morphology of the X-ray emission for both sources

\begin{figure}[!ht]
\centering
\includegraphics[width=0.5\textwidth]{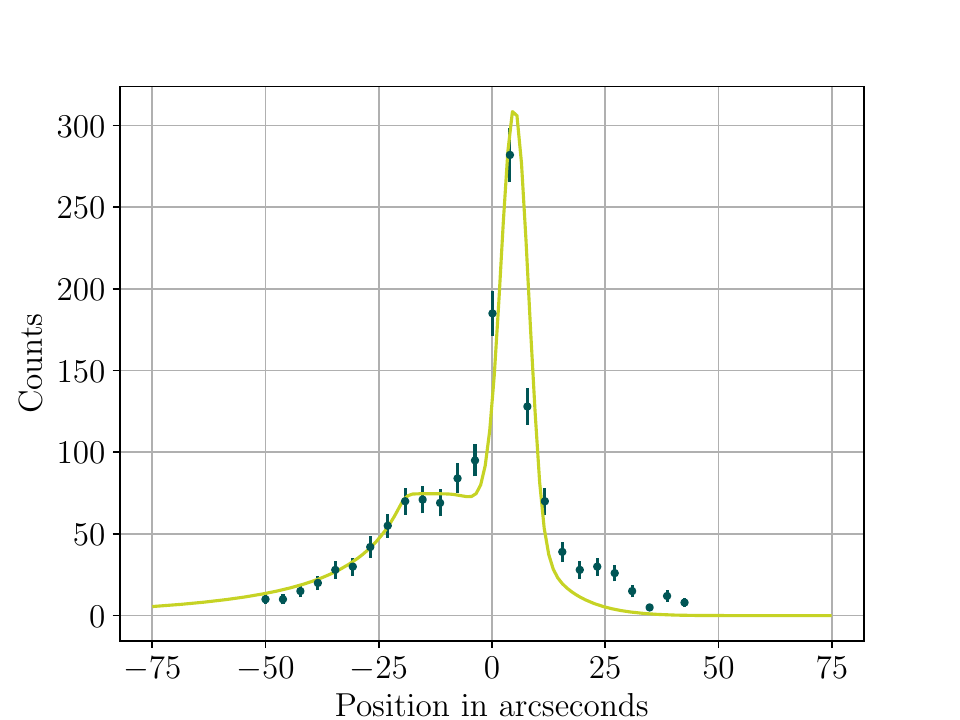}
\caption{
1D profile of our (empirical) analytical model of the X-ray morphology of \hessj\ in comparison to the XMM-Newton measurements from~\cite[Fig. 2]{Funk:2006xk}. 
}
\label{fig:1813_xray_1d_profile}
\end{figure}

\paragraph{\hessj} 
We use an analytical model to describe the morphology of \hessj. The model is based on the $4-7\,\mathrm{keV}$ counts map from  XMM-Newton observations as shown in~\cite[Fig. 2]{Funk:2006xk} and explicitly models the observed morphological asymmetry of the X-ray source. The parameters of this distribution are adjusted so that they match the one-dimensional profile of X-ray counts shown in~\cite[Fig. 2]{Funk:2006xk}. This comparison is shown in Figure~\ref{fig:1813_xray_1d_profile}. The exact parametrisation and parameter values of the model are discussed in detail in Appendix~\ref{app:1813_morphology}.

\begin{figure*}[!ht]
\centering
\includegraphics[trim=0 150 0 150, width=1\textwidth]{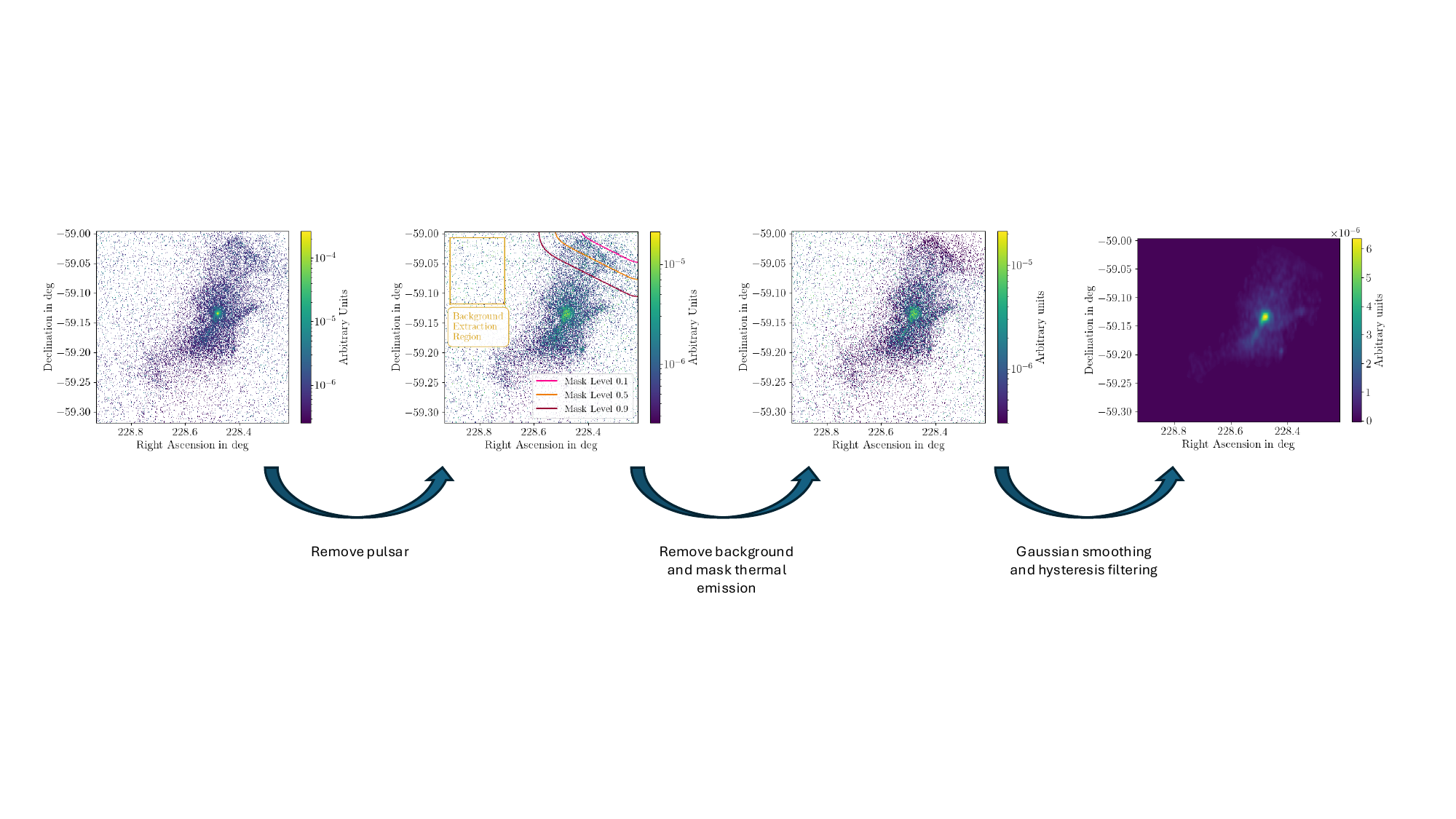}
\caption{
Illustration of the steps to produce a model of \msh\ from the raw, exposure-corrected eROSITA counts map of \msh. In the first step, the emission from the pulsar is removed by replacing the $\approx 100$ brightest pixels close to the pulsar with values sampled from the brightness distribution of the pixels in the two neighboring rows. Then, in the second step, the background is estimated from the area marked by the orange rectangle and removed throughout the image. Furthermore, the thermal-emission from the north-eastern shoulder of \msh\ is masked out with a smoothed mask as indicated by the contours. Finally, the resulting image is smoothed with a $12''$ Gaussian kernel and a hysteresis thresholding is applied.
}
\label{fig:msh_modelling_steps}
\end{figure*}

\paragraph{\msh}For \msh, instead of an analytical model, we directly use X-ray data to create a morphological model. \textcolor{black}{Just like for \hessj, the goal here is not to perform a detailed analysis of the X-ray data, but to create a model template that can be used as input for the TeV studies.} We have opted to use the pointed Early Data Release observations by the eROSITA telescope on board the Spektrum-Roentgen-Gamma (SRG) orbital observatory~\citep{eROSITA:2020emt}. Using the eROSITA Science Analysis Software System (eSASS)~\citep{Brunner:2021hto, eROSITA:2024oyj}, all measured photons with energies between \textcolor{black}{$0.2\,\mathrm{keV}$} and $10\,\mathrm{keV}$ not emitted during flares are selected \textcolor{black}{using the \texttt{evtool} command}. \textcolor{black}{Then, from these events, we select only those with an energy above $2\,\mathrm{keV}$} and produce the raw counts images binned in 8 energy bins of width $1\,\mathrm{keV}$. As this is much wider than the energy resolution of eROSITA~\citep{eROSITA:2020emt}, the effects of energy dispersion can be neglected and these bins are considered bins of true X-ray energy in the following. As the angular pixel size, we choose $4''$. We then divide the counts in each of the bins by the corresponding \textcolor{black}{map of exposure time with vignetting created with eSASS command \texttt{expmap}, multiplied with the on-axis ARF} and sum the bins to generate a single raw flux template map. The steps to produce the final model from this are illustrated in Figure~\ref{fig:msh_modelling_steps}: First, the emission from the pulsar needs to be removed. To do this, we replace the brightness in the $\approx 100$ brightest pixels around the pulsar (corresponding to a radius of $\approx 20''$ around the pulsar) by values randomly sampled from the brightness distribution of the pixels in the two neighboring rows. Then, in order to remove the residual \textcolor{black}{(large-scale)} background, we average the flux measured in a rectangular region away from the source as shown in orange in the second panel in Figure~\ref{fig:msh_modelling_steps} and subtract this value from the map. Even though a large part of the thermal emission from the region is already removed by the chosen $2\,\mathrm{keV}$ energy threshold, we still need to remove the residual thermal emission seen to the north-west of the pulsar. We do this by masking out the corresponding area. The mask is smoothed at the boundary (i.e. not boolean) and then multiplied with the unmasked emission map. Contour lines of this smoothed mask are shown in the second panel of Figure~\ref{fig:msh_modelling_steps}. In total, this mask removes $\approx 10\%$ of the overall emission from the region. While of course the resulting images of the source would appear slightly differently, we therefore expect the precise choice of the mask to only affect the  conclusions of this study at the percent-level. We have verified that even with no mask applied at all, the source model would still be consistent with the current H.E.S.S. results from~\citep{Tsirou:2019kbd}. The mask-multiplied and background-subtracted emission map is shown in the third panel of Figure~\ref{fig:msh_modelling_steps}. The two last steps in the making of the the final morphological map are a gaussian smoothing of the map with a kernel size of $12``$ and cutting out all emission not connected to \msh. The latter is achieved by applying a hysteresis thresholding method. In this, all pixels with a value above a threshold set to the background level are kept if they are connected to the pixel containing the maximum value in the map via pixels also above the threshold. In this way, all islands not connected to \msh\ are discarded. The smoothing (in combination with the energy threshold of $2\,\mathrm{keV}$) also reduces the influence of point sources overlapping with the PWN on the final template morphology. 

Having modelled the morphology of the X-rays, we can now create hypotheses for the spatial distribution of magnetic field and thus high-energy electrons for both sources.

As our baseline assumption we assume a model that has a fixed ratio or equivalently equal spatial distribution between the magnetic field energy density and that of the high-energy electrons, i.e. $n_\mathrm{e}\propto B^2$ along all lines of sight. We call this model the \emph{fixed ratio} model. It is chosen first and foremost as a useful minimal baseline that does not rely on any geometrical prescriptions or scalings for either component. A power-law scaling of the density of high-energy particles with the magnetic field strength is however also expected from MHD theory, with the precise relation depending on the particle transport and magnetic field geometry~\citep{1994plas.conf..225K}.

We then create alternative distributions from this baseline. To do this, we introduce the capping fraction parameter $\eta$ with which we can create capped spatial distributions of either the electron density or the magnetic field energy density from the respective \emph{fixed ratio} distribution as
\begin{equation}
\label{eq:eta_capping}
    f_{\rm cap}=
\begin{cases}
    \eta\times max(f_{\rm fixed}),&f_{\rm fixed}\geq \eta\times max(f_{\rm fixed})\\
    f_{\rm fixed},              & f_{\rm fixed}< \eta\times max(f_{\rm fixed})
\end{cases}
\end{equation}

The corresponding distribution (electrons for capped magnetic energy density and \emph{vice versa}) is then obtained via Equation~\ref{eq:TeV_from_Xray}. Note that this does not mean that the physical maximum values of the respective densities are capped at $\eta\times max(f_{\rm fixed})$, as the distributions are later renormalised to match the total measured gamma-ray source flux.

This parametrization of the alternative hypotheses allows us to create a continuous family of models and to study the sensitivity to differentiate between them as a function of the parameter $\eta$ instead of cherry-picking specific alternative hypotheses. We therefore also choose no preferred value of $\eta$ in the following for either source. The specific values shown in Figures~\ref{fig:1813_model_multipanel} and~\ref{fig:msh_model_multipanel} as well as in Appendix~\ref{app:freepact_maps} are chosen for visualization purposes only.

A capping of the electron density then means that the high intensity X-ray emission from close to the pulsar is attributed to a stronger and more peaked magnetic field in the region. In the opposite case, i.e. a capped magnetic energy density, the electron density is more peaked and concentrated near the pulsar. 

The X-ray morphology together with the morphology of the magnetic field and electron density for the \emph{fixed ratio} model, a model with capped electron density ($\eta=0.5$ for \msh, $\eta=0.25$ for \hessj) and a model with capped magnetic energy density (again $\eta=0.5$ for \msh, $\eta=0.25$ for \hessj) can be seen in Figure~\ref{fig:1813_model_multipanel} for \hessj\ and in Figure~\ref{fig:msh_model_multipanel} for \msh.

\begin{figure*}[!ht]
\centering
\includegraphics[width=1\textwidth]{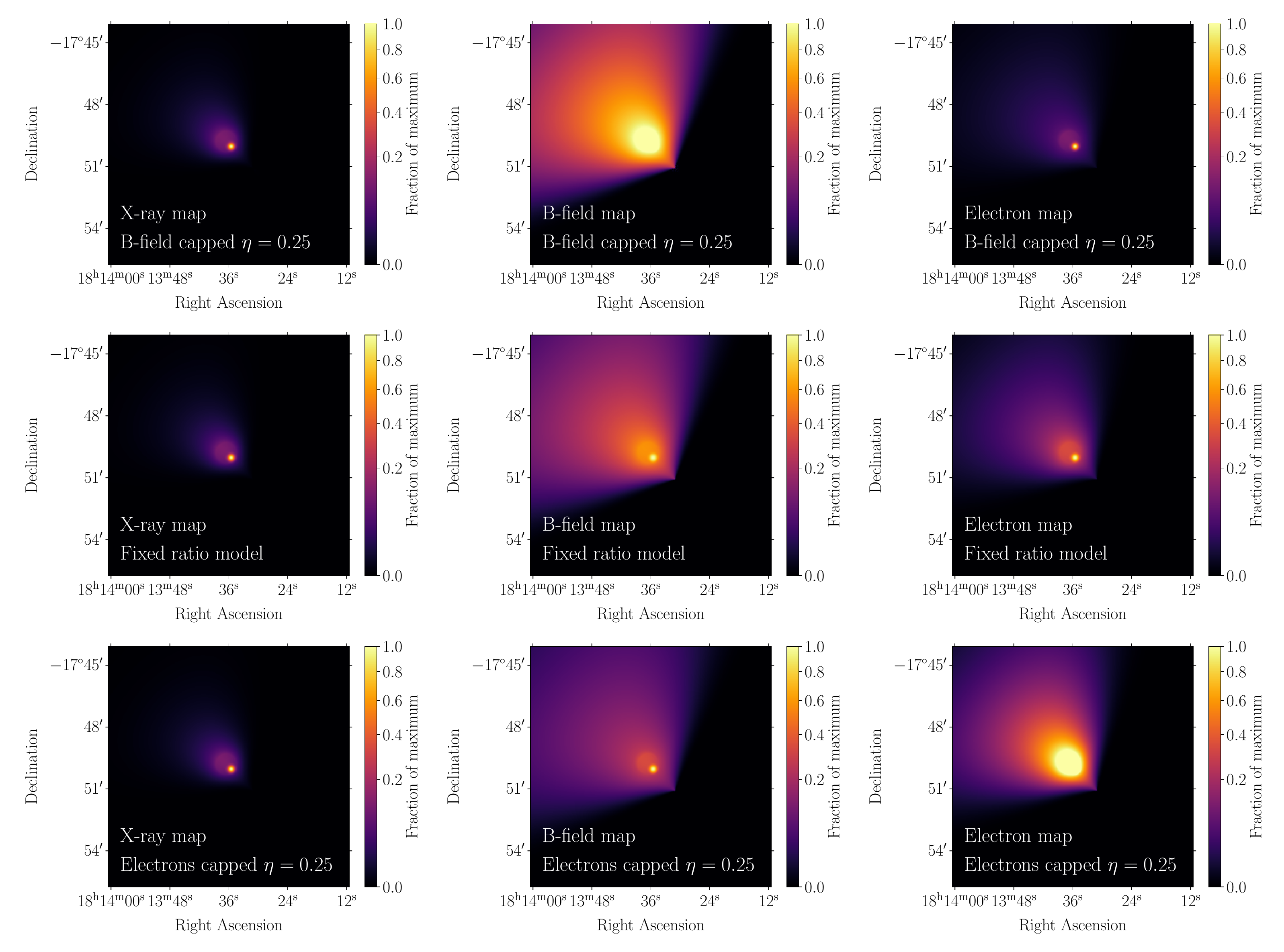}
\caption{
Plot of the different normalised morphological model hypotheses for \hessj. The first column shows the X-ray morphology, the second column the magnetic field and the third column the high-energy electrons for three different models assumptions: The second row shows our baseline \emph{fixed ratio} assumption, in the first row the magnetic field energy density is capped at $\eta=0.25$, in the third row the electron energy density is capped at $\eta=0.25$. Note the square-root color scale.
}
\label{fig:1813_model_multipanel}
\end{figure*}

\begin{figure*}[!ht]
\centering
\includegraphics[width=1\textwidth]{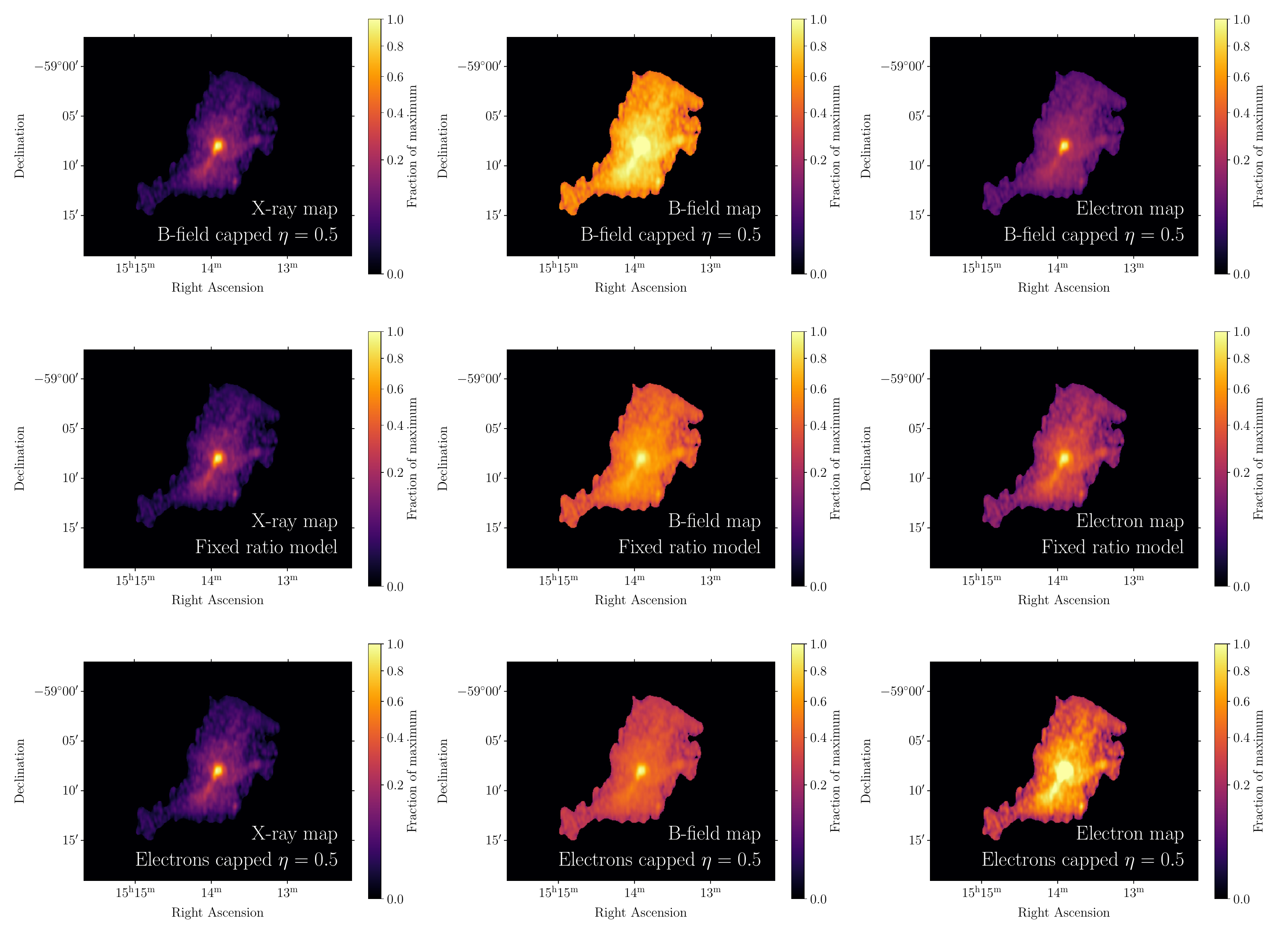}
\caption{
Plot of the different normalised morphological model hypotheses for \msh. The first column shows the X-ray morphology, the second column the magnetic field and the third column the high-energy electrons for three different models assumptions: The second row shows our baseline \emph{fixed ratio} assumption, in the first row the magnetic field energy density is capped at $\eta=0.5$, in the third row the electron energy density is capped at $\eta=0.5$. Note the square-root color scale.
}
\label{fig:msh_model_multipanel}
\end{figure*}

The morphological models obtained in this way are then multiplied by a spectral model and normalized to the total measured source flux. For this, we take the best-fit spectra measured by H.E.S.S. from~\cite{HESS:2024usd} for \hessj\ and from~\cite{Tsirou:2019kbd} for \msh, respectively.

This also means that our modeling approach does not require any quantification of the magnetic field strength or electron density in the sources or a fitting of the measured source spectra with a model for the electron distribution.

However, as a basic cross-check and to validate our approach, we use the \emph{Gamera} tool~\citep{Hahn:2015hhw} to create a simple one-zone time-independent model of the currently radiating electrons and the magnetic field in each source. The details of this model, its fit to data and a discussion of the results can be found in Appendix~\ref{sec:physical_source_fit_method}. Overall, our results from these fits illustrate the broad consistency of our approach with previous models.

\subsection{Creating Fake Datasets}

\begin{figure}[!ht]
\centering
\includegraphics[trim={0cm 1.5cm 0cm 2.9cm},clip,scale=0.49]{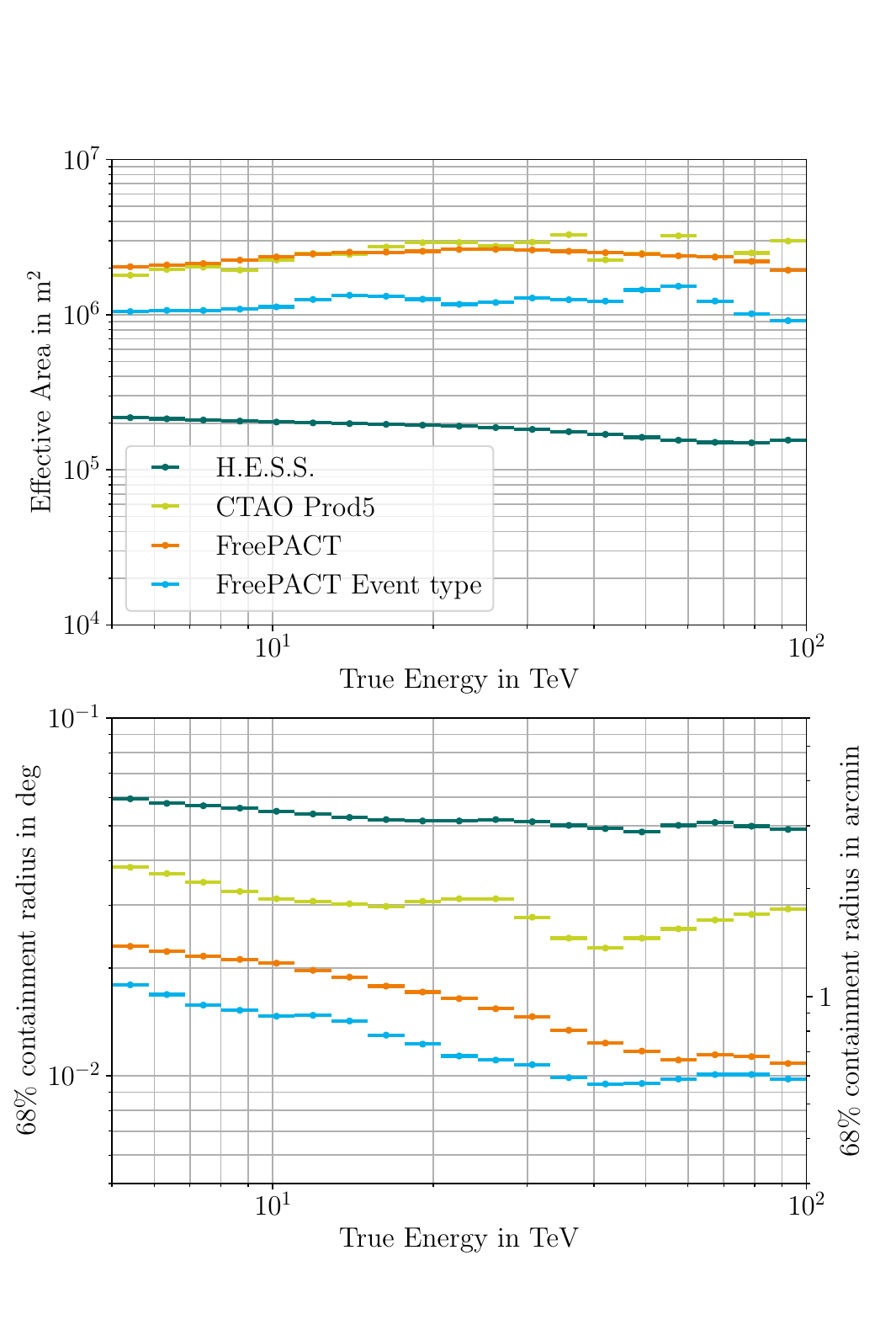}

\caption{
Effective area and angular resolution as a function of true gamma-ray energy for on-axis photons at $20^{\circ}$ zenith angle for the four sets of IRFs used in this work. Shown in dark green are the H.E.S.S. IRFs taken from observation ID 26964 from the public data release~\citep{HESS:2018zix}. In lime green, we show the CTAO \emph{Prod5} IRFs~\citep{cherenkov_telescope_array_observatory_2021_5499840}. The \emph{Free\-PACT} resolution curves are shown in orange. The light blue curve shows the \emph{Free\-PACT}-Event Type IRFs for which only the better half of events in terms of predicted angular resolution are considered. The IRFs are used for the study of \hessj.
}
\label{fig:irf_20deg_1813_errorbars}
\end{figure}

\begin{figure}[!ht]
\centering
\includegraphics[trim={0cm 1.5cm 0cm 2.9cm},clip,scale=0.49]{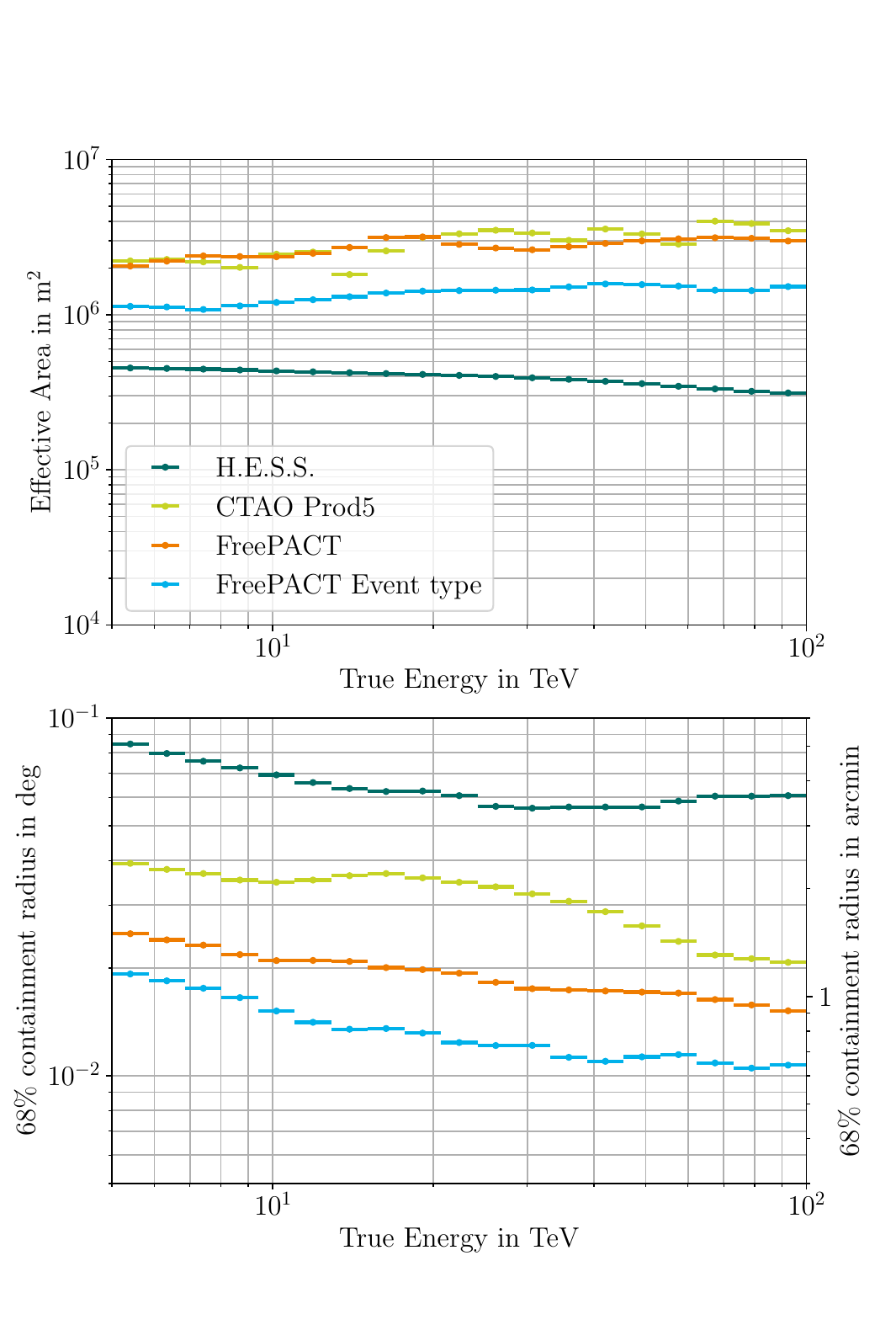}

\caption{
Effective area and angular resolution as a function of true gamma-ray energy for on-axis photons at $40^{\circ}$ zenith angle for the four sets of IRFs used in this work. Shown in dark green are the H.E.S.S. IRFs taken from observation ID 20303 from the public data release~\citep{HESS:2018zix}. In lime green, we show the CTAO \emph{Prod5} IRFs~\citep{cherenkov_telescope_array_observatory_2021_5499840}. The \emph{Free\-PACT} resolution curves are shown in orange. The light blue curve shows the \emph{Free\-PACT}-Event Type IRFs for which only the better half of events in terms of predicted angular resolution are considered. The IRFs are used for the study of \msh.
}
\label{fig:irf_40deg_1813}
\end{figure}

For each of the models described in Section~\ref{sec:source_modelling}, we use version 1.3 of the \emph{gammapy} data analysis package~\citep{gammapy:2023,acero_2023_7734804} to create a set of fake observations characterised by the corresponding Instrument Response Functions (IRFs). In particular, we compare four different sets of IRFs
\begin{itemize}
    \item H.E.S.S.: For this, we use the IRFs made publicly available in the first H.E.S.S. public data release~\citep{HESS:2018zix}. This data release happens to contain observations of \msh. From these, we choose the IRFs associated with observation 20303 in the following. The data release does not contain observations of \hessj. Instead, we choose an observation (26964) of the Kepler SNR which is at a declination similar to \hessj\ to get a comparable set of IRFs. It should be noted that the IRFs provided with this public data release are based on basic reconstruction and gamma-hadron separation algorithms and are not optimized for this analysis. As such, they do not represent the maximum capabilities of H.E.S.S observatory, but rather a baseline performance to compare to.
    
    \item CTAO \emph{Prod5}: We use the IRFs for the \emph{Alpha} configuration of the CTAO southern array determined from the \emph{Prod5} simulation configuration and made available under~\cite{cherenkov_telescope_array_observatory_2021_5499840}. In these, a standard geometrical Hillas method is used for the reconstruction of the source direction~\citep{AHARONIAN1997343}. To simulate observations of \hessj, we use and assume a zenith angle of $20^{\circ}$, for \msh\ we use the IRFs for a zenith angle of $40^{\circ}$.
    
    \item \emph{Free\-PACT}: Similar to the CTAO \emph{Prod5} IRFs, we assume zenith angles of $20^{\circ}$ for \hessj\ and of $40^{\circ}$ for \msh, respectively. The IRFs are based on the processing and reconstruction chain described in~\cite{Schwefer:2024bsx} and in particular make use of the \emph{Free\-PACT} reconstruction algorithm discussed therein.

    This method is part of the family of image-likelihood reconstruction methods that has already been shown to perform well on current-generation instruments~\citep{Parsons:2014voa,deNaurois:2009ud}. In these methods, the shower parameters, both geometry and energy, are reconstructed by maximizing the likelihood of measuring the observed shower images given the shower parameters. Naturally, the formulation of this likelihood function is crucial for the performance of these methods. \emph{Free\-PACT} improves upon previous implementations such as \emph{Im\-PACT}~\citep{Parsons:2014voa} by approximating this likelihood function using the machine-learning based method of \emph{neural ratio estimation} from the field of likelihood-free inference or simulation-based inference~\citep{pmlr-v119-hermans20a,ELLER2023168011}. With this, \emph{Free\-PACT} has been shown to significantly outperform both Hillas-based methods and \emph{Im\-PACT}~\citep{Schwefer:2024bsx}.
    To make it into the final event selection, we require a minimum of two telescopes to pass the set of cuts described in Section 3.4 of~\cite{Schwefer:2024bsx}. As shown in Figures~\ref{fig:irf_20deg_1813_errorbars} and ~\ref{fig:irf_40deg_1813}, this yields effective areas very comparable to the CTAO \emph{Prod5} IRFs. For $20^{\circ}$ zenith angle, we use the same \emph{Free\-PACT} models as in~\cite{Schwefer:2024bsx}, for $40^{\circ}$ zenith angle we train a new set of models. After reconstruction, we apply a Random Forest Classifier trained for gamma-hadron separation to the events. Because optimizing the cut value and estimating the remaining background at multi-TeV energies where background rejection is highly efficient would require a vast number of proton simulations, we simply choose our cut so that $75\%$ of gamma-ray events are kept in each of 10 energy bins. For the background models, we use the corresponding models included in and used for the CTAO \emph{Prod5} IRFs. This is valid because of the similarity between the effective areas of the CTAO \emph{Prod5} IRFs and those of the \emph{Free\-PACT} IRFs. If anything, this estimation of the background for the \emph{Free\-PACT} IRFs is too conservative given that additional information from the \emph{Free\-PACT} fit, such as a goodness-of-fit statistic, could be used in the future.
    
    \item \emph{Free\-PACT}-Event Type: It is foreseen that CTAO will implement \emph{event types}, that is assign each event to a specific class according to its reconstruction quality. While there is no final implementation yet, the approach currently under development~\citep{CTAConsortium:2023ynz} follows that established by Fermi-LAT and sorts events based on the predicted difference between true and reconstructed photon direction. This prediction is done using a neural network algorithm. Here, in order to test the trade-off between effective area (i.e. event statistics) and angular resolution, we use a similar method. For this, we train (for both zenith angles separately) a Random Forest regressor to predict the difference between true and reconstructed source direction. For the training and later evaluation of the regressor, we use a number of reconstructed, event-level parameters, most notably the uncertainty on the direction reconstruction from the \emph{Free\-PACT} fit. For more details on the regressor, see~\ref{app:error_regressor}. We then use this regressor to generate two event types and corresponding IRFs both containing equal amounts of events per energy bin and thus equal effective areas half as large as those of the \emph{Free\-PACT} IRFs. In the following, we only consider the event type featuring the better reconstructed half of events as predicted by the regressor. The corresponding IRFs, as can be seen in Figures~\ref{fig:irf_20deg_1813_errorbars} and~\ref{fig:irf_40deg_1813} will then feature improved angular resolution compared to the \emph{Free\-PACT} IRFs at the cost of halving the effective area. It is not immediately clear whether this is a worthwhile trade-off which makes this an interesting question to investigate. The background model used for the Event type IRFs is the same as for the CTAO \emph{Prod5} and \emph{Free\-PACT} IRFs but - in accordance with the reduction of the effective area - at half the total rate. This is likely a conservative estimate considering that the selection of only well-reconstructed events should enhance the gamma-hadron separation power. Given the large signal-to-background ratio for the observation of these bright compact sources at 10s of TeV (see Tables~\ref{tab:j1813_ncounts} and~\ref{tab:msh_ncounts}), a more precise background estimate would not affect the quantitative conclusions of this study.      
\end{itemize}

For all simulated observations, the source is assumed to be observed on-axis to maximize the angular resolution of the instruments. Because the extensions of the sources and the sizes of the corresponding regions of interest considered here are much smaller than $1^{\circ}$, we neglect the field-of-view offset dependence of the IRFs and consider only their on-axis values (see e.g.~\cite[Fig. 7]{Schwefer:2024bsx} or~\cite{HESS:2006fka} for the validity of this approximation).

All simulated observations only feature one single bin in reconstructed gamma-ray energy ranging from $10\,\mathrm{TeV}$ to $100\,\mathrm{TeV}$. This has two main reasons: First, we are interested only in a morphological (as opposed to a spectral) study of the objects in this work. Then, our models are based on X-ray maps from observations integrated over a rather broad energy band. A finer choice of bins for the gamma-ray maps would feign an energy-dependent modelling that is more detailed and precise than actually present, especially with respect to potential energy-dependent morphology. 

In Figure~\ref{fig:irf_20deg_1813_errorbars}, we compare the effective area and angular resolution for the different IRFs for \hessj, in Figure~\ref{fig:irf_40deg_1813} we show the same for \msh.

\subsection{Testing model separation power}
\label{sec:model_sep_construction}
To statistically ascertain the power of the observations to differentiate between the source models and thus make inferences on the distributions of the magnetic field and high-energy electrons, we use the construction described and applied in~\cite{IceCube:2023hou}. In this, to test the power to differentiate between models $M_1$ and $M_2$, the two test statistic distributions
\begin{equation}
    TS|M_1=-2\log\mathcal{L}(M_2|M_1)+2\log\mathcal{L}(M_1|M_1)
\end{equation}
and
\begin{equation}
    TS|M_2=-2\log\mathcal{L}(M_2|M_2)+2\log\mathcal{L}(M_1|M_2)
\end{equation} are compared. In this, $\log\mathcal{L}(M_i|M_j)$ is the best-fit log-likelihood value obtained from fitting the model $M_i$ to a pseudoexperiment created assuming $M_j$ as the ground truth to sample from. The distributions are then obtained from repeating such pseudoexperiments.

To calculate a figure of merit for the sensitivity to differentiate between the models from these distributions, we choose the two values
\begin{equation}
    p_{\mathrm{ex}}|M_1=p(TS|M_2<\mathrm{med}(TS|M_1))
\end{equation}
and
\begin{equation}
    p_{\mathrm{ex}}|M_2=p(TS|M_1>\mathrm{med}(TS|M_2)).
\end{equation}

In words, $p_{\mathrm{ex}}|M_{1(2)}$ is the median confidence level with which $M_{2(1)}$ can be excluded if $M_{1(2)}$ is true.

\section{Results}
\label{sec:results}

\subsection{Consistency with existing H.E.S.S. measurements}
In order to check the consistency of the models we have produced here with the existing H.E.S.S. measurements of the sources, we fit the same models tested in previous H.E.S.S. analyses of the sources to Asimov datasets. These are mock datasets where the simulated counts exactly equal the model prediction, created from our sky models and the H.E.S.S. IRFs, without any statistical noise~\citep{Cowan:2010js}. The fit is then done exactly as it is done on the actual data in the H.E.S.S. analyses. Using Asimov datasets allows us to eliminate statistical fluctuations from this consistency check and probe the median expected outcome.

For \hessj, we fit a symmetric gaussian spatial model as in~\cite{HESS:2024usd}. We find a source size of $\sigma=0.028^{\circ}$ for the \emph{fixed ratio} model and when capping the electron density at $\eta=0.25$ and $\sigma=0.027^{\circ}$ when capping the magnetic energy density at $\eta=0.25$. This is slightly smaller but consistent with the source size of $\sigma=0.035^{\circ}\pm0.006^{\circ}$ found for energies above $5.7\,\mathrm{TeV}$ in~\cite{HESS:2024usd}. In fact, given that we only consider energies above $10\,\mathrm{TeV}$, a slightly smaller size than the measured one is quite realistic. 

For \msh, we fit an asymmetric gaussian spatial model and compare the sizes of major and minor axis to the values found in~\cite{Tsirou:2019kbd} for energies above $10\,\mathrm{TeV}$, namely $\sigma_{\mathrm{maj}}=0.077^{\circ}\pm0.013^{\circ}$ and $\sigma_{\mathrm{min}}=0.034^{\circ}\pm0.015^{\circ}$. For the major axis size, we again find a somewhat smaller but consistent value of $\sigma_{\mathrm{maj}}=0.059^{\circ}$ for both the \emph{fixed ratio} model and models capping the magnetic or electron energy density at $\eta=0.5$. For the minor axis, we find $\sigma_{\mathrm{min}}=0.032^{\circ}$, in good agreement with the H.E.S.S. measurement. 

\subsection{Counts maps}

\begin{figure*}[!ht]
\centering
\includegraphics[width=1\textwidth]{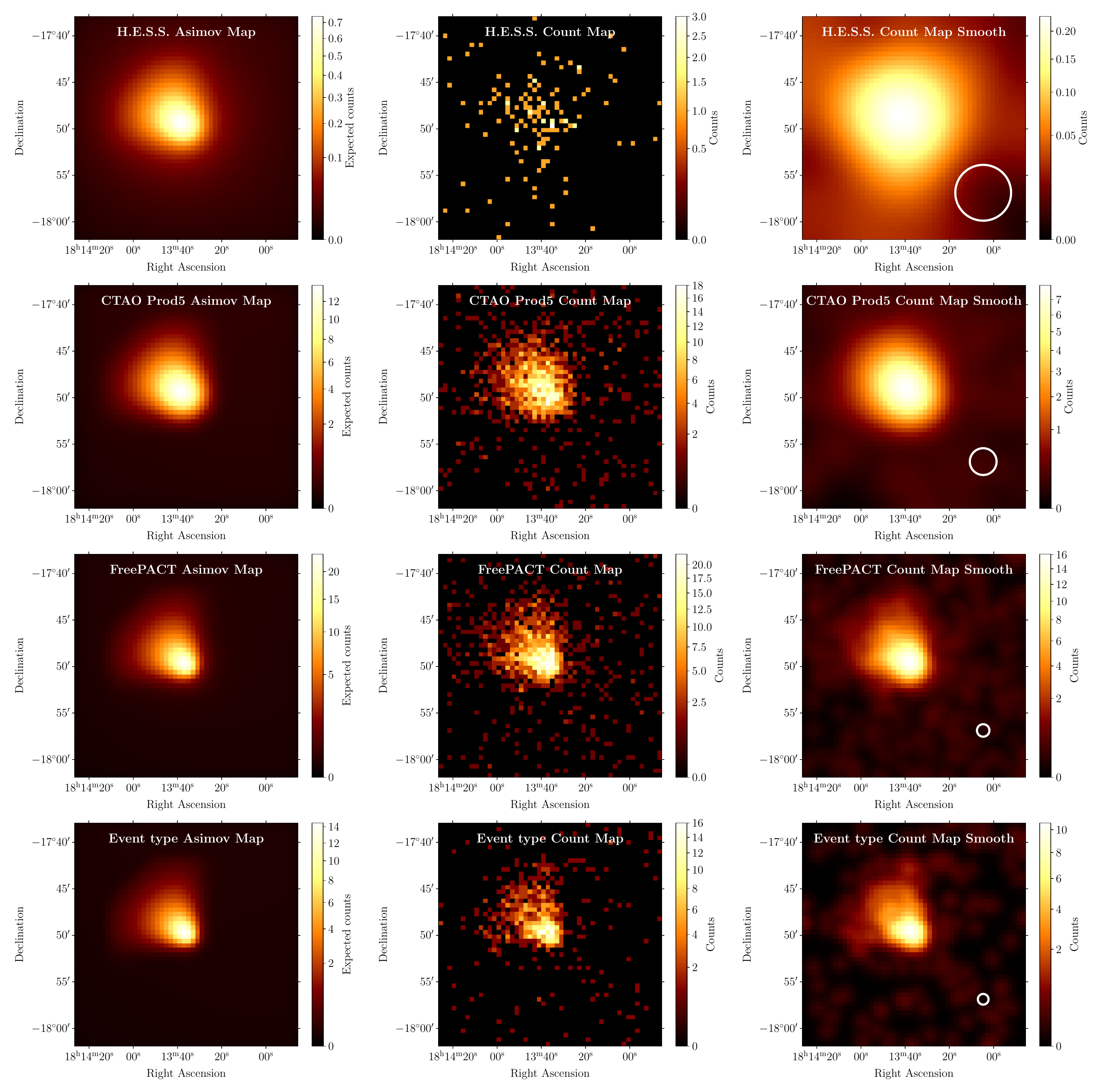}
\caption{
Plot of the expected counts maps for the \emph{fixed ratio} model of \hessj\ for different sets of IRFs between $10\,\mathrm{TeV}$ and $100\,\mathrm{TeV}$ gamma-ray energy. The first column shows the expected counts maps in the infinite-statistics limit (i.e. the Asimov dataset), the second a random realisation for an observation time of $100\,\mathrm{h}$, and the third column shows the map from the second columns smoothed with a gaussian kernel of about the size of the PSF at $50\,\mathrm{TeV}$ indicated by the white circle. The different rows show the different sets of IRFs discussed in the text, which from top to bottom are the H.E.S.S., CTAO \emph{Prod5}, \emph{Free\-PACT} and Event type assumptions, i.e. with increasing resolution from top to bottom. Note the square-root color scales.
}
\label{fig:1813_pred_plot}
\end{figure*}

\begin{figure*}[!ht]
\centering
\includegraphics[width=1\textwidth]{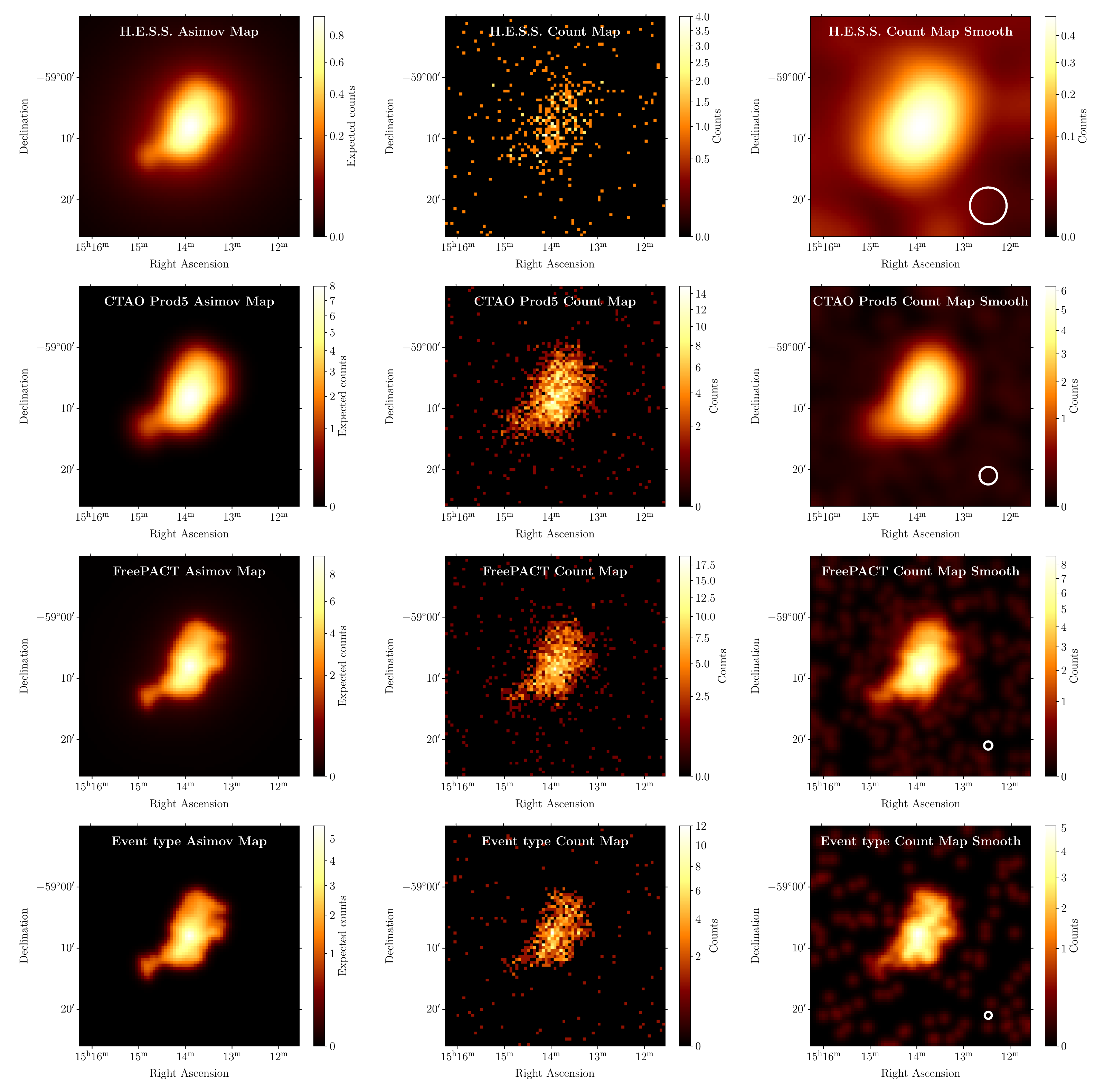}
\caption{
Plot of the expected counts maps for the \emph{fixed ratio} model of \msh\ for different sets of IRFs between $10\,\mathrm{TeV}$ and $100\,\mathrm{TeV}$ gamma-ray energy. The first column shows the expected counts maps in the infinite-statistics limit (i.e. the Asimov dataset), the second a random realisation for an observation time of $100\,\mathrm{h}$, and the third column shows the map from the second columns smoothed with a gaussian kernel of about the size of the PSF at $50\,\mathrm{TeV}$ indicated by the white circle. The different rows show the different sets of IRFs discussed in the text, which from top to bottom are the H.E.S.S., CTAO \emph{Prod5}, \emph{Free\-PACT} and Event type assumptions, i.e. with increasing resolution from top to bottom. Note the square-root color scales.
}
\label{fig:msh_pred_plot}
\end{figure*}

In Figures~\ref{fig:1813_pred_plot} and ~\ref{fig:msh_pred_plot}, we show the dataset counts maps resulting from the convolution of the \emph{fixed ratio} models of \hessj\ and \msh\ with all sets of IRFs. In the first column, we display the expected counts map or Asimov dataset of the source, the second column contains a mock observation counts map for an observation time of $100\,\mathrm{h}$, and the third column shows the mock observation smoothed with a gaussian kernel of approximately the size of the corresponding PSF at $50\,\mathrm{TeV}$.
For \hessj, here and in the following, the simulated datasets also include the extended emission component found in~\cite{HESS:2024usd} with the best-fit parameters given therein.

The rows show the different sets of IRFs discussed in the text, which from top to bottom are the H.E.S.S., CTAO \emph{Prod5}, \emph{Free\-PACT} and Event type assumptions, i.e. with increasing resolution from top to bottom.

The corresponding total number of expected signal and background events are given in Tables~\ref{tab:j1813_ncounts} and~\ref{tab:msh_ncounts}.Table~\ref{tab:j1813_ncounts} also include the predicted counts of the extended emission component.

A figure showing smoothed counts map for the \emph{Free\-PACT} IRFs and different model assumptions for both sources that visualize the model separation power that is quantitatively discussed in Section~\ref{sec:model_separation_result} can be found in Appendix~\ref{app:freepact_maps}. 

In all cases, the count maps alone are enough to clearly identify the sources, it is not necessary to compute a significance map for this purpose. This is in line with the signal-to-background ratios that can be read off from Tables ~\ref{tab:j1813_ncounts} and~\ref{tab:msh_ncounts}, which range from $\approx 3$ for H.E.S.S. on \msh\ to $\approx 16$ for the CTAO observations of \hessj. Clearly, the residual hadron background is not the limiting factor to the studies presented here.

Clearly visible in both figures is the increased effective area and thus event statistics that will be possible with the CTAO southern array at these energies. This is owed mostly to the 37 SSTs that will be spread out over an area much larger than covered by the H.E.S.S. telescopes. In numbers, the source events are increased by a factor $13$ for \hessj\ and a factor $6$ for \msh\ for the CTAO \emph{Prod5} IRFs compared to H.E.S.S..

Looking at the first column of both figures, it appears that resolving the asymmetry of \hessj\ and the features of \msh\ is possible even with H.E.S.S. at these energies given infinite observation time. However, looking at the second and third column of both figures shows that this would require a lot more than the already fairly generous $100\,\mathrm{h}$ of observation time assumed here. 

Even with the CTAO \emph{Prod5} IRFs, it is difficult to resolve the asymmetry of \hessj. For \msh, the source is clearly elongated, but making out the jet in the south-east of the source (see Section~\ref{sec:introduction} and~\citep{Gaensler:2001ac} for a detailed description of this feature) is already difficult. Still, it is a clear improvement over the capabilities of H.E.S.S..

With the \emph{Free\-PACT} and Event type IRFs, these features become clearly visible and even the finger-like filaments in the north-east of \msh\ (see~\ref{sec:introduction} and~\citep{Gaensler:2001ac}) start to appear. This is of course no quantitatively significant statement, but illustrates the improvements and progress that improved reconstruction algorithms promise to bring to the CTAO. 

\begin{table}
\centering
\label{tab:j1813_ncounts}
\caption{Expected count numbers for \hessj\ signal and background between $10\,\mathrm{TeV}$ and $100\,\mathrm{TeV}$ in $100\,\mathrm{h}$ of observations for the different sets of IRFs. The third column contains the counts predicted for the extended emission component found in~\cite{HESS:2024usd}.}
\begin{tabular}{c|c|c|c} IRFs & \makecell{Counts\\ \hessj} & \makecell{Counts\\ Extended} & Backgr.\\ \hline 
H.E.S.S.& 100&15&14 \\
CTAO \emph{Prod5} & 1313&196&80\\
\emph{Free\-PACT} & 1317&203& 80\\
Event type & 668&105&40\\
\end{tabular}
\end{table}

\begin{table}
\centering
\label{tab:msh_ncounts}
\caption{Expected count numbers for \msh\ signal and background between $10\,\mathrm{TeV}$ and $100\,\mathrm{TeV}$ in $100\,\mathrm{h}$ of observations for the different sets of IRFs.}
\begin{tabular}{c|c|c} IRFs & \makecell{Counts\\ \msh} & Background\\ \hline 
H.E.S.S.& 312& 98\\
CTAO \emph{Prod5} & 1872&190\\
\emph{Free\-PACT} & 1679&190\\
Event type & 803&95\\
\end{tabular}
\end{table}

\subsection{Model separation}
\label{sec:model_separation_result}
To test the model separation power, we use the statistical construction described in Section~\ref{sec:model_sep_construction}. To limit the number of tests, we always choose $M_1$ to be the respective \emph{fixed ratio} model and then test the separation power relative to models with different values of the capping parameter $\eta<1$ (see Equation~\ref{eq:eta_capping}) for either the electron or the magnetic field energy density as $M_2$. 

For each value of $\eta$, we simulate 50000 mock observations for each source and set of IRFs. From these, we obtain the distributions of $TS|M_1$ and $TS|M_2$ and then calculate the values of $p_{\mathrm{ex}}|M_1$ and $p_{\mathrm{ex}}|M_2$ in two different ways: in case there is enough overlap between the distributions, we determine them empirically from the simulated samples. If not, we fit both distributions with gaussians and determine the overlap from the fitted curves. An example of this is shown in Appendix~\ref{app:ts_distribution}. Finally, we convert the p-values to gaussian significance.

For all cases, the considered simulated observation time is $100\,\mathrm{h}$.

\begin{figure}[!ht]
\centering
\includegraphics[clip,scale=0.49]{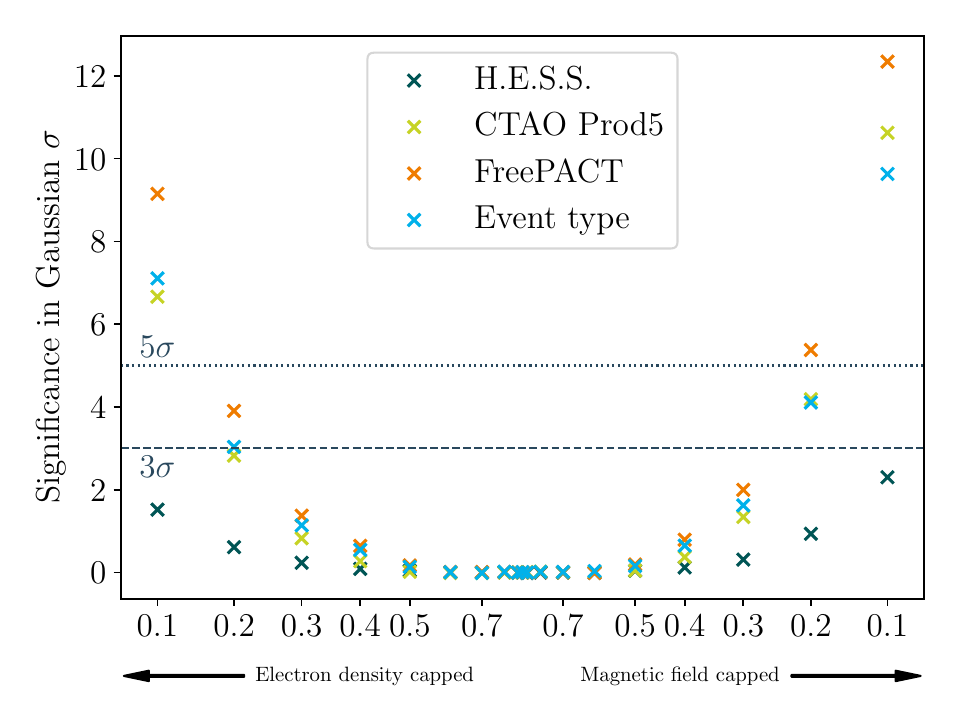}

\caption{
Median expected significance for the separation of the \emph{fixed ratio} model to the capped models of \hessj\ for different values of the cap fraction $\eta$, assuming the \emph{fixed ratio} model to be true.
}
\label{fig:sensitivities_1813}
\end{figure}

\begin{figure}[!ht]
\centering
\includegraphics[clip,scale=0.49]{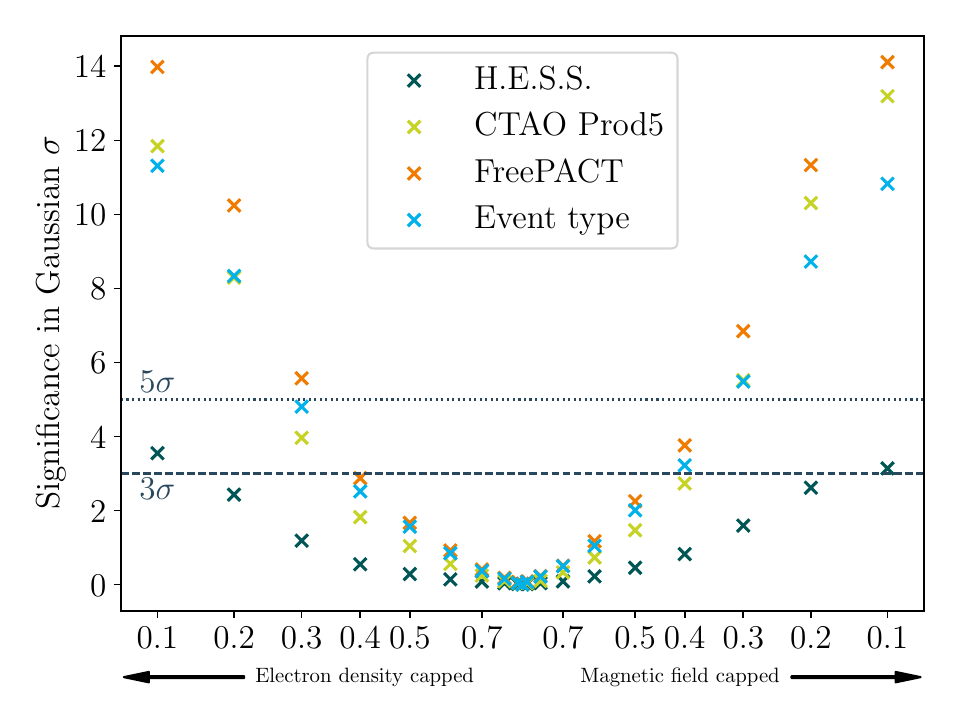}

\caption{
Median expected significance for the separation of the \emph{fixed ratio} model to the capped models of \msh\ for different values of the cap fraction $\eta$, assuming the \emph{fixed ratio} model to be true.
}
\label{fig:sensitivities_msh}
\end{figure}

The resulting sensitivities for \hessj\ are shown in Figure \ref{fig:sensitivities_1813}, for \msh\ they are shown in Figure \ref{fig:sensitivities_msh}. These Figures only show the $p_{\mathrm{ex}}|M_1$ case, i.e. the scenario in which the \emph{fixed ratio} model is assumed to be the ground truth. For the opposite case, where the respective capped model is assumed as the ground truth, see~\ref{app:sensitivity_reversed}. While quantitative values may differ slightly between the two cases, the qualitative results are the same.

There are several conclusions that can be drawn: First, as expected, the capability to differentiate between the models in any case dramatically improves with the increased sensitivity of CTAO compared to H.E.S.S.. Also, as expected, as the value of $\eta$ decreases in either direction, the ability to differentiate the capped model from the \emph{fixed ratio} model increases correspondingly.

A direct comparison of the two sources is difficult because of the specific definition of $\eta$ relative to the maximum of each source. The qualitative conclusion that the significances for the model separation in both sources are in the same order of magnitude with a slight advantage towards \msh\ is not surprising given the similar extensions and total brightness of the sources, with slightly more events expected from \msh.

Regarding the impact of the improved angular resolution achievable with the \emph{Free\-PACT} reconstruction, it indeed provides a measurable boost in sensitivity for the study of these sources relative to the CTAO \emph{Prod5} IRFs featuring a very similar effective area. For example, for an electron capping at $\eta=0.2$ for \hessj, it increases the median expected significance from $2.8\,\sigma$  to $3.9\,\sigma$, when capping the magnetic field energy density at $\eta=0.2$ the increase is from $4.2\,\sigma$ to $5.3\,\sigma$. Put differently, the chances of a $5\,\sigma$ significant differentiation between the models increases from $0.8\%$ to $16\%$ in case of the capped electron density and from $21\%$ to $66\%$ for the capped magnetic field energy density.

Considering the performance of our Event type IRFs, it becomes clear that for these sources, the trade-off of improved angular resolution whilst sacrificing effective area, is not advantageous. In all scenarios considered, the Event type IRFs perform significantly worse than the \emph{Free\-PACT} IRFs and on par with the CTAO \emph{Prod5} IRFs. This shows that in the energy range above $10\,\mathrm{TeV}$ considered here, event statistics are still a significant limitation to the characterisation of source morphology. Therefore, a simple increase of the observation time remains one of the easiest ways to reveal more about the nature of sources, knowing that \textcolor{black}{it will be in high demand} at facilities such as the CTAO.

The potential of the Event type approach also goes beyond what is presented here in a couple of ways: First, we have only considered statistical uncertainties in this study. However, systematic uncertainties will also play a role in a realistic study and eventually - with increasing observation time - come to dominate the uncertainty. In that case, the increased statistics of observations assuming the \emph{Free\-PACT} IRFs compared to the Event type IRFs considered here will not reduce the overall uncertainty, and the improved angular resolution of the Event type IRFs will gain value correspondingly.

And then, for this study, we have only looked at the first of two event types, featuring the better-reconstructed half of events, to test the trade-off between angular resolution and effective area. However, when used in a joint analysis with the other half of events and its set of IRFs, it will always perform at least as well as the regular \emph{Free\-PACT} IRFs based on the same events but only featuring one set of IRFs. Therefore, our findings should by no means be seen as an argument against the introduction of event types. 

Finally, a lowering of the energy threshold could significantly increase the available statistics. Even going as low as $5\,\mathrm{TeV}$ could potentially double the event statistics with a PSF still better than for the CTAO \emph{Prod5} IRFs. Here, we do not consider this option further, for one to keep consistency between the methods but first and foremost to stay within the limits set by our analysis approach requiring a one-to-one correspondence between the X-ray and gamma-ray morphologies.
We also note that the lower-energy data is not without value for constraining the source properties for any of the IRFs and that it can be incorporated into the analysis in any case using a more complex physical source-modeling approach~\citep{Egg:2025cdc}.

\section{Conclusion}
\label{sec:conclusion}
In this paper, we presented the case for PWNe as targets for future high-energy, high-precision measurements with CTAO and especially its southern array.

PWNe are frequent and bright sources at multi-TeV energies and feature plentiful structure at arcminute scales known from X-ray observations. These structures contain information on the physical properties of the sources, in particular the distribution of electrons and magnetic field.

The southern array of the CTAO, in particular in combination with improved gamma-ray reconstruction methods such as \emph{Free\-PACT}, will be able to reach arcminute resolution at energies above $10\,\mathrm{TeV}$.

Using \hessj\ and \msh\ as examples, we have demonstrated that these high-resolution measurements in combination with existing X-ray measurements allow for much enhanced insights into the distributions of high-energy particles and magnetic fields of the sources.

To this end, we have developed  different models of the TeV morphology of \hessj\ and \msh\ based on X-ray observations and consistent with physical models and previous H.E.S.S. measurements of the sources. We have then calculated the sensitivity to differentiate between these models and constrain the distributions of electrons and magnetic fields assuming different sets of observations and corresponding IRFs from H.E.S.S. and CTAO.

As expected, we find CTAO to significantly increase the sensitivity of these observations relative to H.E.S.S.. The sensitivity of CTAO can itself be significantly increased through improved reconstruction algorithms such as \emph{Free\-PACT}. However, we find that sacrificing event statistics in favor of even greater angular resolution to not be a worthwhile trade-off, suggesting that event statistics are still a significant limitation to these multi-TeV observations.

For future studies of the TeV morphology of PWNe and the physical insights that can be gained from them, it will be interesting to go beyond the simple empirical modeling approach followed here and create more physically motivated models of the sources. Such models, by explicitly modelling e.g. the interplay between particle transport and magnetic field geometry, the injection of high-energy particles into the nebula, their energy losses and the history of the pulsar, can overcome the limitation of the approach followed here set by the coarse matching of the energy ranges between the observed X-ray and gamma-ray data throughout the entire source. Such models can also extend to lower energies and thus help to harness the information contained in the energy-dependent morphology of the sources which we have neglected here by focusing on a rather narrow energy range.   
The conclusions from the present study however still apply even for such more complex models. Different model assumptions will still produce similar X-ray signatures due to the inherent degeneracy between particles and magnetic fields, and, as demonstrated here, high-resolution observations with the CTAO provide an opportunity to differentiate between them through the observed gamma-ray morphology.

Via Equation~\ref{eq:TeV_radiation}, the morphology of the TeV emission itself also contains a degeneracy between the density of high-energy particles and the intensity of the target radiation field for Inverse Compton scattering. Here, we work under the assumption of a homogeneous target radiation field comprising the large-scale ISRF and the CMB. This assumption may not be valid in particular source environments and a more careful modeling of the radiation field structure might be required in such cases (see~\cite{Conte:2024mgw} for an example for the case of the Galactic Centre region). However, having produced such models for the particular intensity distribution of the target radiation, they can be applied in Equation~\ref{eq:TeV_radiation}, and all further calculations and conclusions drawn here still apply when replacing $I_{\mathrm{TeV}}\rightarrow I_{\mathrm{TeV}}/n_{\mathrm{rad}}$. The effects of a varying spectrum of the radiation field throughout the source on the gamma-ray spectrum are expected to be small because the scattering of infrared photons occurs in the Klein-Nishina regime for the electron energies considered here (see Section~\ref{sec:source_modelling}). Because of the inherent non-linearity of the process, synchrotron self-Compton scattering is difficult to treat in the framework used here. However, it is only expected to contribute significantly for very few particularly young and powerful systems such as the Crab nebula~\citep{Torres:2013jha}. For the case of \msh, it has been explicitly shown that the contribution of this process is negligible~\citep{Tsirou:2019kbd}.

The analysis presented here also still leaves significant room for improvement and optimization for the multi-TeV energy range and the high-resolution study of small sources: For one, in case of limited event statistics as present in the scenarios considered here, it could become feasible to use an unbinned per-event likelihood formalism as is commonly employed for example in searches for astrophysical neutrinos~\citep{IceCube:2022der}. This would in particular replace the PSF averaged over all events with individual per-event PSFs, potentially enhancing the resolution and significance of the observation. 
The analysis can be further optimized for the multi-TeV energy range through a dedicated event selection procedure optimized on this energy range only. This could for example result in less stringent gamma-hadron cuts, allowing more background but also increasing the number of signal events.
Finally, there is also the option to employ the observatory in convergent pointing mode. This reduces the field of view of the observatory but increases the average telescope multiplicity of events. This is desirable for the on-axis, high-resolution study of small sources as presented here, as the smaller field of view \textcolor{black}{is not relevant} and the higher multiplicity leads to better angular resolution. This was studied in more detail for CTAO in~\cite{Szanecki:2015zaa}.

The advent of modular open-source data analysis frameworks such as \emph{gammapy} has now also opened up the possibility of performing direct, likelihood-level, joint analyses of gamma-ray and X-ray data~\citep{Giunti:2022edz,NievasRosillo:2024xwp,Egg:2025cdc}. In this context,  the specific case of a joint analysis of \msh\ from eROSITA and H.E.S.S. data has been recognized as particularly interesting and will be discussed in a forthcoming publication~(see~\cite{Egg:2025cdc}). Analyses of this type will allow to fully harness the physical information contained in the multiwavelength measurement of the sources.

In this study, we have presented the high-resolution study of PWNe with CTAO. There are however more source classes whose study is likely to profit from the improved angular resolution of CTAO:

SNRs are typically of larger angular size, but also feature observable structure on arcminute scales, known from X-ray observations or - in case of hadronic emission - arising due to the structure of the target gas. One prominent candidate source the morphology of which has already been studied extensively is RX J1713.7-3946. For example, a recent study by H.E.S.S.~\citep{refId0} probes the radial profile of the TeV emission in comparison to X-ray measurements of the source and finds high-energy particles leaking beyond the shock front seen in X-rays. There is also a sensitivity study considering the potential of the CTAO to differentiate between leptonic emission (tracing the X-rays) and hadronic emission (tracing the gas) in RX J1713.7-3946~\citep{CTAConsortium:2017xgl}. Both of these studies are also likely to profit from the improved angular resolution achievable with \emph{Free\-PACT}, demonstrating the potential for high-resolution measurements of SNRs with the CTAO.

Another class of TeV gamma-ray sources that has gotten a lot of attention recently and has been shown to emit gamma rays up to hundreds of TeV are microquasars~\citep{LHAASO:2024psv,doi:10.1126/science.adi2048}. Clearly, the improved exposure and resolution of the CTAO will be just as beneficial to detailed spectro-morphological studies such as the H.E.S.S. analysis of SS 433~\citep{doi:10.1126/science.adi2048} as it is to the studies of PWNe discussed here.

Finally, while absorption on the extragalactic background light limits the TeV brightness of these sources, nearby extragalactic sources could also be interesting targets for high-resolution studies with CTAO. For example, resolving the jet structure of Centaurus A beyond the recent findings of H.E.S.S.~\citep{Abdalla:2020tlj} could shine new light on the acceleration processes and distributions of relativistic particles in these sources.

\section*{Acknowledgements}

 This work was conducted in the context of the CTAO Consortium. We thank Katharina Egg for invaluable help with obtaining and processing the eROSITA data of MSH 15-52. We are furthermore grateful to Katharina Egg, Alison Mitchell, Fabio Acero and Dan Parsons for giving valuable feedback that significantly improved the manuscript. G.S. acknowledges membership in the International Max Planck Research School for Astronomy and Cosmic Physics at the University of Heidelberg (IMPRS-HD).

This work made use of Gammapy \citep{gammapy:2023}, a community-developed Python package. The Gammapy team acknowledges all Gammapy past and current contributors, as well as all contributors of the main Gammapy dependency libraries: \href{https://numpy.org/}{NumPy}, \href{https://scipy.org/}{SciPy}, \href{http://www.astropy.org}{Astropy}, \href{https://astropy-regions.readthedocs.io/}{Astropy Regions}, \href{https://scikit-hep.org/iminuit/}{iminuit}, \href{https://matplotlib.org/}{Matplotlib}.

This research has made use of the CTA instrument response functions provided by the CTA Consortium and Observatory, see https://www.ctao-observatory.org/science/cta-performance/ (version prod5 v0.1; \cite{cherenkov_telescope_array_observatory_2021_5499840}) for more details.

This work made use of data from the H.E.S.S. DL3 public test data release 1 (HESS DL3 DR1, H.E.S.S. collaboration, 2018).

This work is based on data from eROSITA, the soft X-ray instrument aboard SRG, a joint Russian-German science mission supported by the Russian Space Agency (Roskosmos), in the interests of the Russian Academy of Sciences represented by its Space Research Institute (IKI), and the Deutsches Zentrum für Luft- und Raumfahrt (DLR). The SRG spacecraft was built by Lavochkin Association (NPOL) and its subcontractors, and is operated by NPOL with support from the Max Planck Institute for Extraterrestrial Physics (MPE). The development and construction of the eROSITA X-ray instrument was led by MPE, with contributions from the Dr. Karl Remeis Observatory Bamberg \& ECAP (FAU Erlangen-Nuernberg), the University of Hamburg Observatory, the Leibniz Institute for Astrophysics Potsdam (AIP), and the Institute for Astronomy and Astrophysics of the University of Tübingen, with the support of DLR and the Max Planck Society. The Argelander Institute for Astronomy of the University of Bonn and the Ludwig Maximilians Universität Munich also participated in the science preparation for eROSITA.  The eROSITA data shown here were processed using the eSASS software system developed by the German eROSITA consortium. 
\bibliographystyle{aa} 
\bibliography{my_library.bib}

\begin{appendix}

\section{Analytical model of \hessj}
\label{app:1813_morphology}
For \hessj, the morphology of the X-ray intensity is modelled using the following analytical distribution:
\begin{equation}
\begin{split}
    I(x,y)&=F(x,y,x_{0,F},y_{0,F},r_{0},\alpha,F_{+},F_{-})\times\\
    &(N_1\times\mathcal{G}_1(x,y,x_{0,1},y_{0,1},\sigma_{x,1},\sigma_{y,1})\\
    &+N_2\times\mathcal{G}_2(x,y,x_{0,2},y_{0,2},\sigma_{x,2},\sigma_{y,2})).
\end{split}
\end{equation}

In this, $x$ and $y$ denote the directions along the major and minor axis of the source, respectively. We take the major axis to be oriented in a north-east to south-west direction as used for the extraction of the 1D emission profile in~\cite[Fig. 2]{Funk:2006xk}.
Defining $r = \sqrt{(x - x_{0,F})^2 + (y - y_{0,F})^2}$, the distribution $F(r)$ is chosen as 
\begin{equation}
     F(r) = \max\left(\begin{cases} F_{\text{+}}, & \text{if } r < r_0 \\ F_{\text{+}} \left(\frac{r}{r_0}\right)^{-\alpha}, & \text{if } r > r_0 \end{cases}, F_{-}\right).
\end{equation}
$\mathcal{G}_1$ and $\mathcal{G}_2$ are gaussian distributions. $\mathcal{G}_1$ is chosen to be symmetric, whereas $\mathcal{G}_2$ is asymmetric with
\begin{equation}
\sigma_{y,2}(x)=\begin{cases} s_{\text{m}}(x_{\text{m}} - x), & \text{if } s_{\text{m}}(x_{\text{m}} - x) > 0 \\ 0, & \text{otherwise} \end{cases}    
\end{equation}
and
\begin{equation}
    \sigma_{x,2}(x) = \begin{cases} \sigma_{\text{n},1}, & \text{if } x \leq x_0 \\ \sigma_{\text{n},2}, & \text{if } x > x_0 \end{cases} 
\end{equation}
depending on $x$.
\begin{table}
\centering
\label{tab:j1813_param_values}
\caption{Parameter values for the analytical model of the X-ray morphology of \hessj.}
\begin{tabular}{c|c} Parameter & Value \\ \hline $x_{0,F}$ & -20'' \\
$y_{0,F}$ & 0'' \\
$r_{0}$ & 30'' \\
$F_{\text{+}}/F_{\text{-}}$ & 100 \\
$\alpha$ & 1.4 \\
$N_{1}/N_{2}$ & 8 \\
$x_{0,1}$ & 10'' \\
$y_{0,1}$ & 0'' \\
$\sigma_{x,1}=\sigma_{y,1}$ & 6'' \\
$x_{0,2}$ & -30'' \\
$y_{0,2}$ & 0'' \\
$x_{\text{m}}$ & 100'' \\
$s_{\text{m}}$ & 0.33 \\
$\sigma_{\text{n},1}$ & 135'' \\
$\sigma_{\text{n},2}$ & 40'' \\
\end{tabular}
\end{table}

The adopted parameter values are given in Table~\ref{tab:j1813_param_values}. These are tuned to reproduced the measured 1d X-ray emission morphology of \hessj\ as given in ~\cite[Fig. 2]{Funk:2006xk}. Since the overall normalisation of the model is arbitrary at this point and later determined to reproduce the total source flux, only the ratios of the normalisation parameters are listed in Table~\ref{tab:j1813_param_values}.

\section{Physical source parameters}
\label{sec:physical_source_fit_method}

Here, we briefly describe our simple one-zone time-independent model of the currently radiating electrons and the magnetic field in each source that we use to cross-check our approach with previous models of the sources. For this, we use the \emph{Gamera} tool~\citep{Hahn:2015hhw}. Note that the parameters determined in this fit are not required for the descriptions in the main text and purely calculated for validation purposes.

For these models, the spatial distributions of both components are entirely fixed through our hypotheses created in Section~\ref{sec:source_modelling}. The only free parameters are the overall normalisations of the magnetic field and electron distributions and the spectral index and high-energy cutoff of the electron spectrum modeled as a cut-off power-law
\begin{equation}
    n_{e}(E)=NE^{-\alpha}\exp{(-E/E_{\text{cut}})}.
\end{equation}

The approach to determine these parameters is then as follows: We first calculate the parameters of the electron spectrum and distribution from a fit to H.E.S.S. data from both sources (taken from~\cite{HESS:2024usd} for \hessj\ and from~\cite{Tsirou:2019kbd} for \msh). Then we use X-ray data (from XMM-Newton~\citep{Funk:2006xk} for \hessj\ and from BeppoSAX~\citep{Mineo:2001fr} for \msh) to fix the only free parameter left, the overall normalisation of the magnetic field.

As the minimum electron energy in this we choose $100\,\mathrm{GeV}$. This ensures the entire H.E.S.S. spectrum can be used for the parameter determination. It is below the energy range of electrons for which our spatial model is valid as discussed above. However, since we integrate over the full source for the comparison to H.E.S.S. data, this procedure is independent of the morphology within. For the comparison to X-ray data, we take care to only integrate over those regions of our models also used for the extraction of the fluxes by the different instruments. As we only consider X-ray energies above $1\,\mathrm{keV}$, only multi-TeV electrons contribute to the flux, validating the usage of our spatial model.

As the target radiation field for the inverse Compton scattering, we assume the CMB together with the ISRF model from~\cite{Popescu:2017dgo} evaluated at the respective source positions. We assume \hessj\ to be at a distance of $6.2\,\mathrm{kpc}$~\citep{Camilo:2021jmy} and \msh\ to be at a distance of $5.2\,\mathrm{kpc}$~\citep{Gaensler:1999xf}.

For both sources, we use three different morphological models in the fit: The \emph{fixed ratio} model and one model capping the electron density and the magnetic field energy density each, at $\eta=0.25$ for \hessj\ and at $\eta=0.5$ for \msh.

\begin{table*}
\centering
\caption{Best-fit peak line-of-sight averaged magnetic field strength $B_\mathrm{max}$ for the fits to the three tested models for \hessj\ and \msh.}
\label{tab:magnetic_field_vals}
\begin{tabular}{c|c|c|c} Parameter & \emph{fixed ratio} model & \makecell{Electrons capped\\ \hessj: $\eta=0.25$\\ \msh: $\eta=0.5$}& \makecell{Magnetic field capped\\ \hessj: $\eta=0.25$\\ \msh: $\eta=0.5$}\\ \hline 
\hessj& $16\,\mathrm{\mu G}$&$31\,\mathrm{\mu G}$&$8\,\mathrm{\mu G}$ \\
\msh& $18\,\mathrm{\mu G}$&$25\,\mathrm{\mu G}$&$13\,\mathrm{\mu G}$ \\
\end{tabular}
\end{table*}

\begin{figure}[!ht]
\centering
\includegraphics[width=0.5\textwidth]{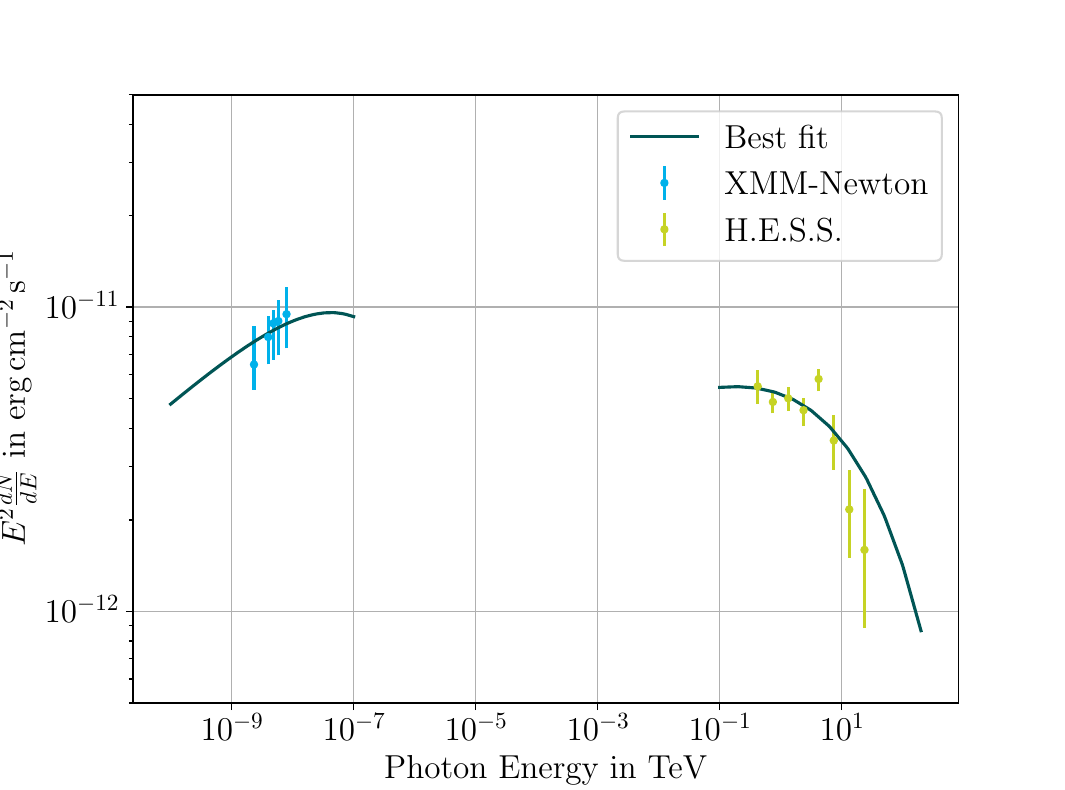}
\caption{
Spectral energy distribution for \hessj\ from XMM-Newton and H.E.S.S, observations, showing the best fit synchrotron and inverse Compton spectra from a one-zone time-independent model. For this Figure, the \emph{fixed ratio} model was assumed. See text for more details.
}
\label{fig:multiwvl_sed_1813.pdf}
\end{figure}

\begin{figure}[!ht]
\centering
\includegraphics[width=0.5\textwidth]{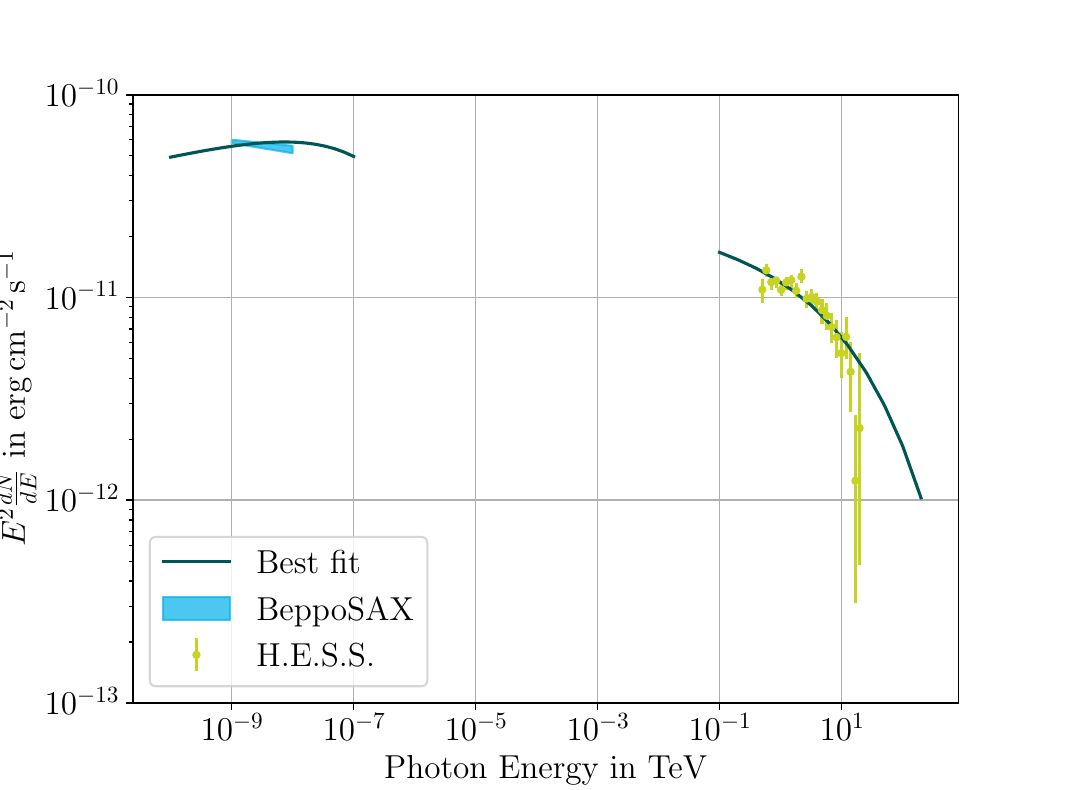}
\caption{
Spectral energy distribution for \msh\ from BeppoSAX and H.E.S.S, observations, showing the best fit synchrotron and inverse Compton spectra from a one-zone time-independent model. For this Figure, the \emph{fixed ratio} model was assumed. See text for more details.
}
\label{fig:multiwvl_sed_msh.pdf}
\end{figure}

The resulting best-fit photon spectra are shown in Figures~\ref{fig:multiwvl_sed_1813.pdf} and~\ref{fig:multiwvl_sed_msh.pdf}.

The best-fit parameters related to the electron distribution are the same for all morphological models for a given source. That is because these concern only the full-source integrated flux and spectrum of the source. For \hessj, we find the total energy in electrons with energies $>1\,\mathrm{TeV}$, $W_{E>1\,\mathrm{TeV}}$, to be  $3\times10^{47}\,\mathrm{erg}$ and a spectral index of $\alpha=2.65$. For \msh, the corresponding values are $W_{E>1\,\mathrm{TeV}}=7\times10^{47}\,\mathrm{erg}$ and $\alpha=2.86$.

The high-energy cutoff of the electron spectrum $E_{\rm cut}$ is not determined from the data, but fixed to $1\,\mathrm{PeV}$ as it could not be constrained in the fit. We have verified that the values of the other parameters are not significantly affected by the choice of the cutoff.

The single-powerlaw assumption for the spectrum hides the fact that the electron spectra in these sources should exhibit a cooling break and thus softening towards higher energies. The values found for the spectral index $\alpha$ therefore interpolate between the cooled and uncooled part of the electron spectrum. Keeping this in mind, our results for $\alpha$ are in line with what would be expected from a single powerlaw fit of the spectra~\citep{Giacinti:2019nbu}. They are also broadly consistent  with the values found for these and other PWN sources in other analyses (e.g.~\cite{Liu:2024jdt, HESS:2024usd, Tsirou:2019kbd}).

To ascertain the plausibility of the values found for $W_{E>1\,\mathrm{TeV}}$, we compare it to two values for each source: An estimate of the total available from the pulsar spin-down given by $\dot{E}\tau_\mathrm{c}$ and the values found for the total energy in electrons above $1\,\mathrm{GeV}$ in~\cite{Liu:2024jdt}. For the total spin-down energy, we find values of $10^{49}\,\mathrm{erg}$ for \hessj\ and $10^{48}\,\mathrm{erg}$ for \msh. The values from~\cite{Liu:2024jdt} are $1.6\times10^{48}\,\mathrm{erg}$ for \hessj and $1.9\times10^{48}\,\mathrm{erg}$ for \msh. Our values for $W_{E>1\,\mathrm{TeV}}$ are broadly consistent with these: As should be the case, they are smaller than our reference values, but also suggest that a significant fraction of the total energy goes into the TeV leptons and thus that the PWNe are efficient particle accelerators.

Regarding the strength of the magnetic field, we quote the peak line-of-sight averaged magnetic field $B_{\rm max}$ in our model in Table~\ref{tab:magnetic_field_vals}. Here, we see a clear effect when changing the morphology of the electrons/magnetic field distributions: As expected, when the magnetic field energy density is capped, we find a lower peak magnetic field in the source and \emph{vice versa} for the case of a capped electron density.

The best-fit values on the order of $\approx 10-20\,\mathrm{\mu G}$ are very much in line with previous estimates for both sources (e.g~\cite{HESS:2024usd,HESS:2005lqd,Gaensler:2001ac}) and other PWNe~\citep{Liu:2024jdt}.

It should be noted that both the values of the total energy in electrons and the peak magnetic field are strongly degenerate with the intensity of the IRSF at the source. A stronger ISRF would require less high-energy electrons to produce the measured gamma rays and therefore a stronger magnetic field to produce the X-rays. As discussed in~\cite{Breuhaus:2020mof}, an enhancement of a factor 2 is certainly plausible in spiral arm regions or near discrete sources. Such an enhancement would lead to an increase of the magnetic field of the order $\sqrt2$.

Overall, the parameter values obtained in these fits illustrate the broad consistency of our approach with previous models.

\section{Angular error regressor}
\label{app:error_regressor}

To train the Random Forest regressor designed to predict the difference between true and reconstructed gamma-ray direction, the following reconstructed event-level parameters are used:
\begin{itemize}
    \item Uncertainty of the reconstructed gamma-ray source direction as determined from the \emph{Free\-PACT} fit
    \item Reconstructed gamma-ray energy as determined from the \emph{Free\-PACT} fit
    \item Relative uncertainty of the reconstructed gamma-ray energy as determined from the \emph{Free\-PACT} fit.
    \item Reconstructed distance of the shower core location on ground from the center of the array as determined from the \emph{Free\-PACT} fit
    \item Uncertainty of the reconstructed distance of the shower core location on ground from the center of the array as determined from the \emph{Free\-PACT} fit
    \item Number of triggered telescopes
    \item Number of telescopes passing the event selection criteria as described in Section 3.4 of~\cite{Schwefer:2024bsx}.
\end{itemize}

For each zenith angle, a regressor is trained on 100000 events ranging in energy from $20\,\mathrm{GeV}$ to $200\,\mathrm{TeV}$. Note that this also means that the regressor is not specifically optimized for the multi-TeV energy range.

\begin{figure}[!ht]
\centering
\includegraphics[trim={0cm 3cm 0cm 4cm},clip,scale=0.46]{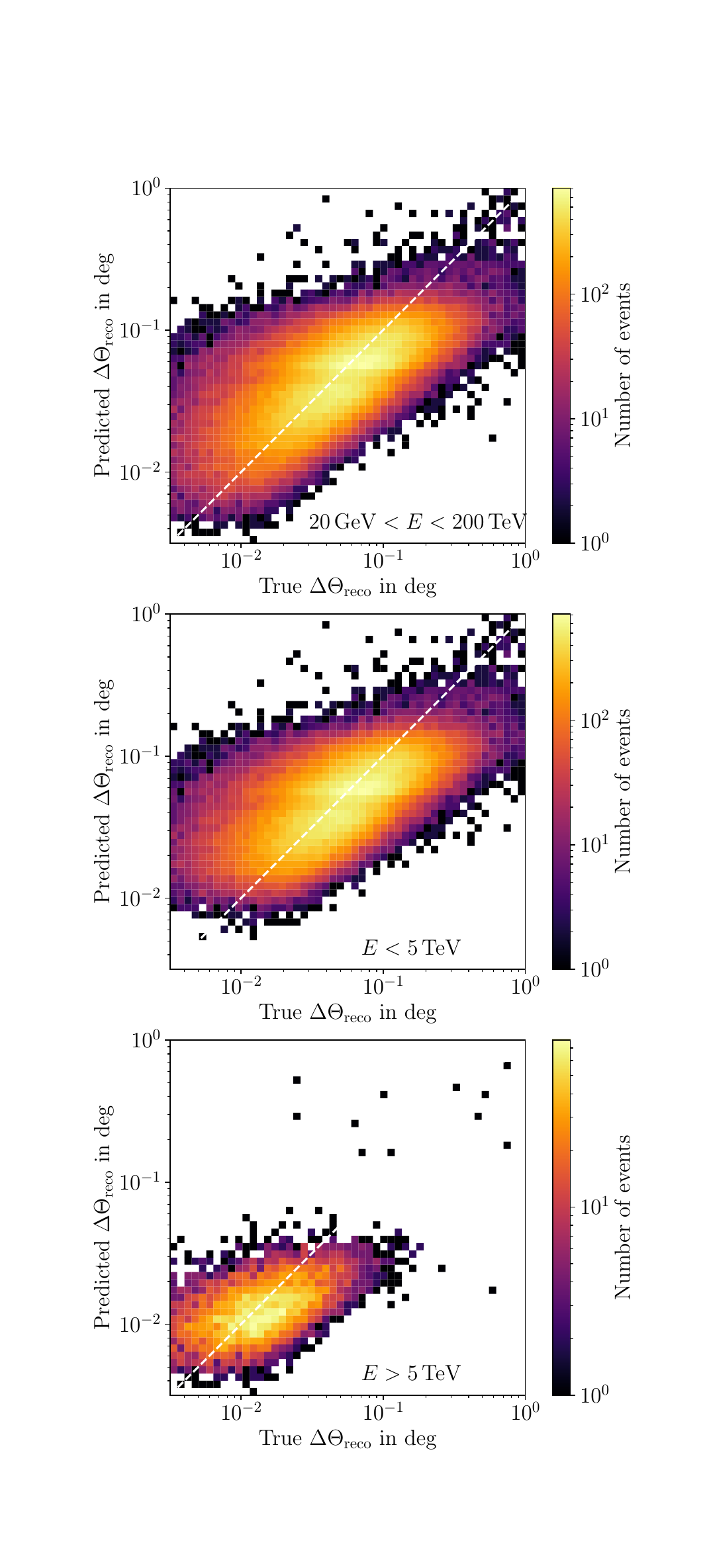}

\caption{
Plot of true vs. reconstructed difference between true and reconstructed gamma-ray source direction as predicted by the Random Forest regressor for a zenith angle of $20^{\circ}$.
}
\label{fig:event_type_regressor}
\end{figure}

Figure~\ref{fig:event_type_regressor} illustrates the performance of the regressor on a test data set for a zenith angle of $20^{\circ}$, averaged over all energies and below and above $5\,\mathrm{TeV}$. As can be seen, despite not being specifically optimized for multi-TeV energies, it still performs adequately in that energy range.

\section{Test statistic distribution}
\label{app:ts_distribution}

\begin{figure}[!ht]
\centering
\includegraphics[clip,scale=0.47]{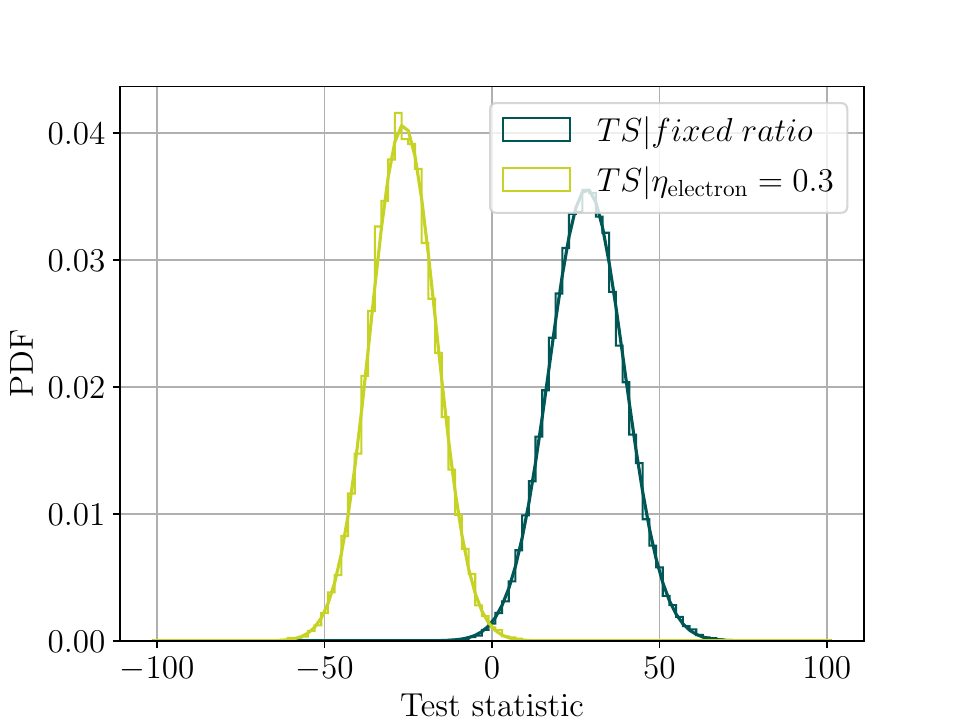}

\caption{
Distributions of the test statistics $TS|M_1$ and $TS|M_2$ and gaussian fits for \msh\ assuming the \emph{Free\-PACT} IRFs. We assume the \emph{fixed ratio} model as $M_1$ and a model with the electron distribution capped at $\eta=0.3$ as $M_2$. The distributions are generated from 50000 mock observations assuming $100\,\mathrm{h}$.
}
\label{fig:ts_distributions}
\end{figure}

Here we show as an example distributions of the test statistics $TS|M_1$ and $TS|M_2$ for \msh\ assuming the \emph{Free\-PACT} IRFs. We assume the \emph{fixed ratio} model as $M_1$ and a model with the electron distribution capped at $\eta=0.3$ as $M_2$. The distributions are generated from 50000 mock observations assuming $100\,\mathrm{h}$. In this case, since neither distribution has samples beyond the median of the other, we use fitted gaussian distributions to determine $p_{\mathrm{ex}}|M_1$ and $p_{\mathrm{ex}}|M_2$. As can be seen in Figure~\ref{fig:ts_distributions}, the gaussian distributions are a good description of the distribution from pseudo-experiments.

\section{Count maps for \emph{FreePACT} IRFs for different model hypotheses}
\label{app:freepact_maps}

Here, we show smoothed count maps above $10\,\mathrm{TeV}$ of different models for \hessj\ and \msh\ assuming the \emph{Free\-PACT} IRFs. All mock observations assume and observation time of $100\,\mathrm{h}$ and the maps are smoothed with a 0.7 arcminute disk kernel. These figures visualize the power to separate different model hypotheses with enhanced angular resolution that is quantitatively discussed in Section~\ref{sec:model_separation_result}.

\begin{figure*}[!ht]
\centering
\includegraphics[clip,scale=0.33]{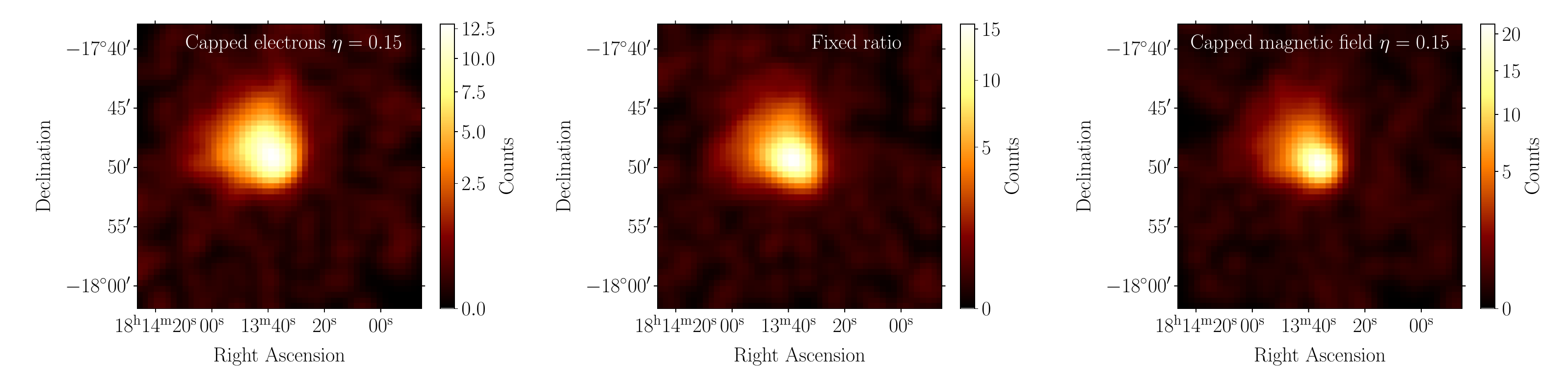}

\caption{
Smoothed count maps for mock observations of \hessj, assuming the \emph{FreePACT} IRFs. From left to right, the three windows assume a model of \hessj\ with the electron density capped at $\eta=0.15$, the \emph{fixed ratio} model and a model with the magnetic field energy density at $\eta=0.15$. Note the square-root color scales.
}
\label{fig:freepact_1813_model_count_maps}
\end{figure*}

\begin{figure*}[!ht]
\centering
\includegraphics[clip,scale=0.33]{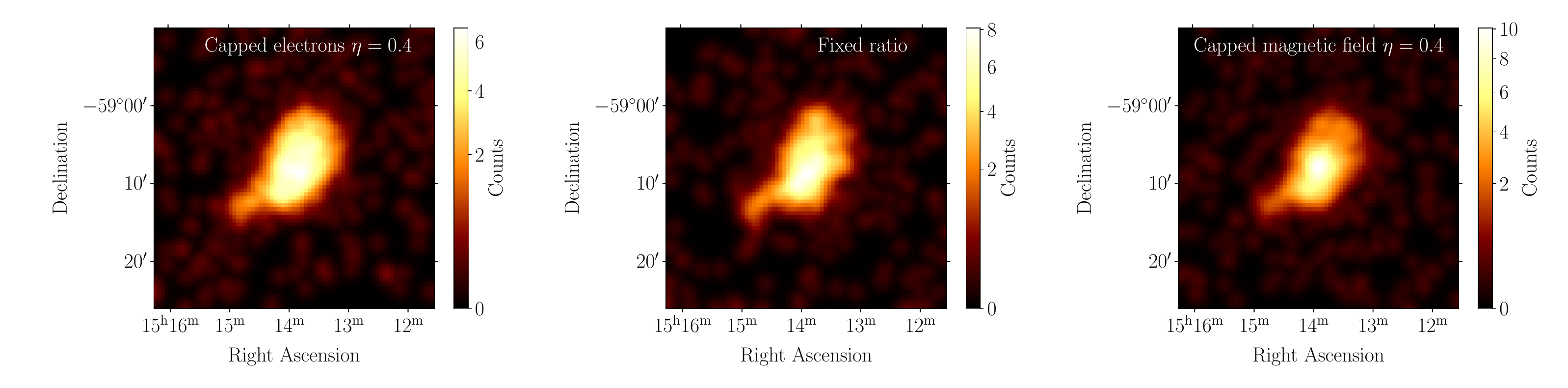}

\caption{
Smoothed count maps for mock observations of \msh, assuming the \emph{FreePACT} IRFs. From left to right, the three windows assume a model of \hessj\ with the electron density capped at $\eta=0.4$, the \emph{fixed ratio} model and a model with the magnetic field energy density at $\eta=0.4$. Note the square-root color scales.
}
\label{fig:freepact_msh_model_count_maps}
\end{figure*}

\section{Sensitivity relative to capped models}
In Section~\ref{sec:model_separation_result}, we only show the sensitivity to differentiate between the \emph{fixed ratio} model and capped models for both sources \textit{assuming the fixed ratio model to be true}. Here, we show the opposite case, where the respective capped model is assumed to be true. While the precise values might be different to the case described in Section~\ref{sec:model_separation_result}, all qualitative conclusions there also apply here.
\label{app:sensitivity_reversed}
\begin{figure}[!ht]
\centering
\includegraphics[clip,scale=0.49]{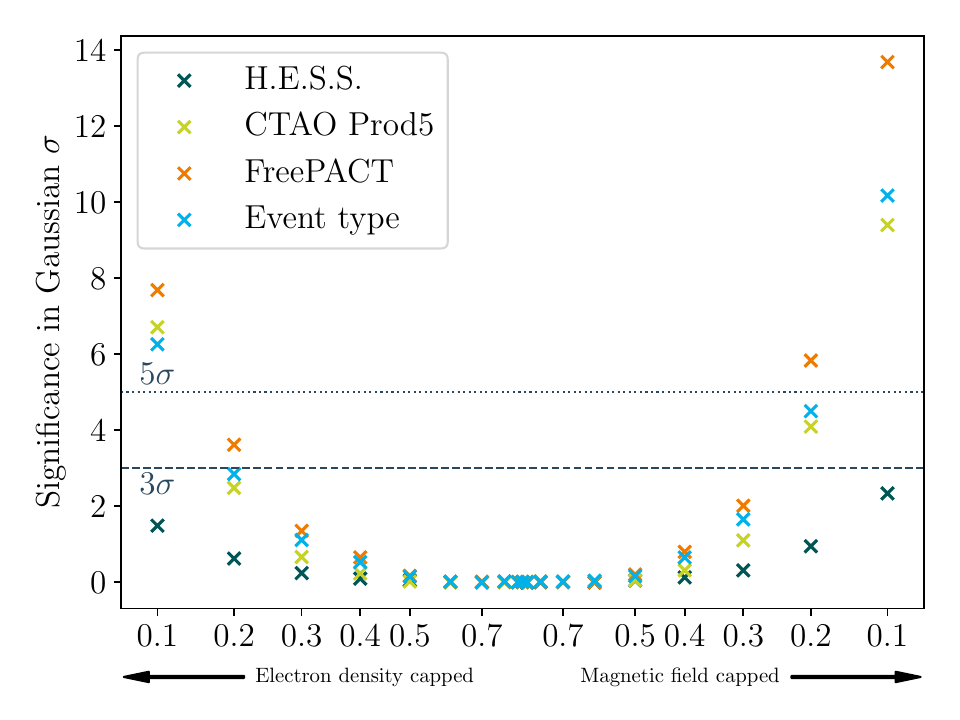}

\caption{
Median expected sensitivity for the separation of the \emph{fixed ratio} model to the capped models of \hessj\ for different values of the cap fraction $\eta$, assuming the respective capped model to be true.
}
\label{fig:sensitivities_1813_reverse}
\end{figure}

\begin{figure}[!ht]
\centering
\includegraphics[clip,scale=0.49]{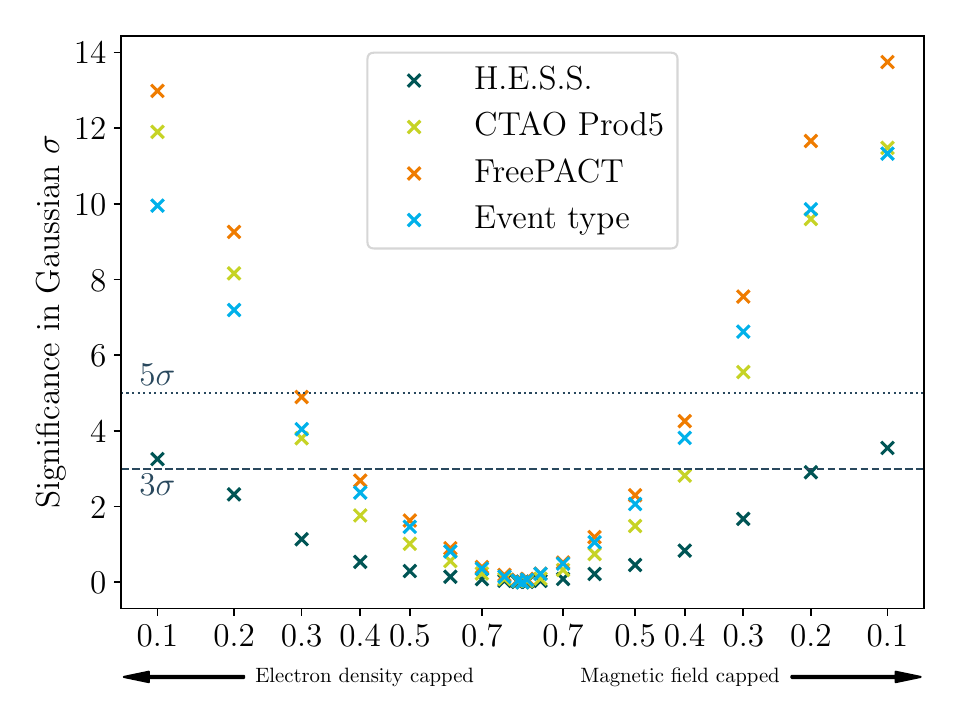}

\caption{
Median expected sensitivity for the separation of the \emph{fixed ratio} model to the capped models of \msh\ for different values of the cap fraction $\eta$, assuming the respective capped model to be true.
}
\label{fig:sensitivities_msh_reverse}
\end{figure}

\end{appendix}
\end{document}